\documentclass[10pt,review]{siamltex}
\usepackage{ifthen,algorithm,algorithmic}
\usepackage{amssymb,amsfonts,amsmath,latexsym,dsfont}
\usepackage{tikz,pgfplots,pgflibraryplotmarks}
\usetikzlibrary{calc}
\usepackage{graphicx}
\usepackage{mathrsfs}
\usepackage{algorithm,algorithmic}
\usepackage{hyperref}

\graphicspath{{images/}{./}}

\usepackage{boxedminipage}

\usepackage{pgf,pgfarrows} 
\usepackage{subfigure} 

\newcommand{\FrameboxA}[2][]{#2}
\newcommand{\Framebox}[1][]{\FrameboxA}

\newcommand{\hf}{{\frac 12}}

\newcommand{\jInv}{{\sf jInv}}

\renewcommand{\div}{\nabla\cdot\,}

\newcommand{\bfA}{{\bf A}}

\newcommand{\bfF}{{\bf F}}

\newcommand{\bfJ}{{\bf J}}

\newcommand{\bfM}{{\bf M}}

\newcommand{\bfu}{{\bf u}}
\newcommand{\bfq}{{\bf q}}

\newcommand{\bfd}{{\bf d}}
\newcommand{\bfm}{{\bf m}}

\newcommand{\bfw}{{\bf w}}
\newcommand{\bfv}{{\bf v}}

\newcommand{\bfp}{{\bf p}}

\newdimen\iwidth
\newdimen\iheight

\newcommand{\CA}{{\cal A}}

\renewcommand{\theequation}{\arabic{section}.\arabic{equation}}

\newcommand{\R}{\ensuremath{\mathds{R}}}
\newcommand{\C}{\ensuremath{\mathds{C}}}

\newtheorem{example}{Example}

\newtheorem{remark}{Remark}

\newsavebox{\theFAIRBox}
 {\end{minipage}\end{lrbox}%
  \pagebreak[2]\vskip5pt\noindent\fcolorbox{yellow!20}{yellow!20}{\box\theFAIRBox}\vskip2mm}

\newcommand{\RTHedit}[1]{{#1}}

\begin{document}
\title{jInv -- a flexible Julia package for PDE parameter estimation}
\author{Lars Ruthotto\thanks{Department of Mathematics and Computer Science, Emory University, Atlanta, Georgia, USA. ({\tt lruthotto@emory.edu})} \and Eran Treister\thanks{Department of Computer Science, Ben-Gurion University of the Negev, Beer Sheva, Israel. ({\tt erant@cs.bgu.ac.il})} \and Eldad Haber\thanks{ Department of Earth and Ocean Sciences, University of British Columbia, Vancouver, Canada. ({\tt haber@math.ubc.ca.})}}
\maketitle
\begin{abstract}
	Estimating parameters of Partial Differential Equations
	(PDEs) from noisy and indirect measurements often requires solving ill-posed inverse problems. These so called parameter estimation or inverse medium problems arise in a variety of applications such as geophysical, medical imaging, and nondestructive testing.
	 Their solution is computationally intense since the underlying PDEs need to be solved numerous times until the reconstruction of the parameters is sufficiently accurate.
	 Typically, the computational demand grows significantly when more measurements are available, which poses severe challenges to inversion algorithms as measurement devices become more powerful.

	In this paper we present jInv, a flexible framework and open source software that provides parallel algorithms  for solving parameter estimation problems with many measurements.
	Being written in the expressive programming language Julia, jInv is portable, easy to understand and extend, cross-platform tested, and well-documented.
	It provides novel parallelization schemes that exploit the inherent structure of many parameter estimation problems and can be used to solve multiphysics inversion problems as is demonstrated using numerical experiments motivated by geophysical imaging.
\end{abstract}



\section{Introduction}

Many inverse problems can be formulated as parameter estimation problems that involve Partial Differential
Equations (PDEs) as forward problems.
Problems of this kind arise in many applications.
Our main focus in this paper is geophysical imaging, where parameter estimation is used, e.g., in Direct Current (DC) resistivity~\cite{deymor}, electromagnetic
inverse problems~\cite{haberBook2014,schulz2011computational}, Full Waveform Inversion (FWI)~\cite{pratt1999,EpanomeritakisAkcelikGhattasBielak2008,krebs09ffw,VirieuxOperto2009,WarnerEtAt2013}, history matching~\cite{OliverBook2008}, and travel time tomography~\cite{sei1994gradient,leung2006adjoint,taillandier2009first,li2013first}.
Structurally similar inverse problems arise in medical imaging, e.g., in Diffuse Optical Tomography~~\cite{Arridge1999, ArridgeSchotland2009,SaibabaEtAl2015} or Electrical Impedance Tomography~\cite{CheneyEtAl1999}.

In the above applications, the inverse problem consists of estimating physical properties of an object, e.g., the electric conductivity or the seismic velocity of the subsurface from surface measurements. Typically, data is recorded for a number of experiments that can be characterized  by the location, geometry and frequency of the source and similar information about the receivers. The forward problems then consist of simulating the data for a given model and involve solving the governing PDEs for all experiments. During a typical inversion the forward problems have to be solved several times until the estimate of the model is sufficiently accurate, thus requiring a large number of PDE solves.

Over recent years, data collection techniques have been constantly improving and thus inversion problems involving data from many sources, receivers and frequencies need to be solved more commonly.
When the experiments are independent of one another, a parallel implementation of the forward problem and, in the context of inversion,
a parallel implementation of optimization techniques is desirable.
In particular, codes that scale well in the number of sources and frequencies are of major interest.

While implementing parallel inversion codes is simple in principle, it can be hard in practice. There are two main
obstacles when attempting to implement a parallel framework for inverse problem.
First, from an algorithmic point of view, many classical parallel implementations do not exploit the structure of PDE parameter estimation problems, in which the forward problems require non-trivial amounts of memory. Second, writing efficient parallel code required
(until recently) using lower level computing languages and was difficult to modify and work with.
Recently, a new high-level programming language, Julia, with strong focus on parallel scientific computation was developed~\cite{BezansonEtAl2012}.
Julia enables rapid prototyping by using an expressive syntax that is similar to MATLAB or Python.
Julia includes a Just-In-Time (JIT) compiler that produces code with similar efficiency than that of C or Fortran compilers.
Most importantly in our context, Julia provides easy to use tools for parallel computing.
This makes this language an excellent platform for the solution of
PDE parameter estimation problems, especially for problems that require parallelization.

\RTHedit{In this paper we present \jInv, which is a free and open source software and framework that allows users
to easily implement forward problems, set up inverse problems and solve them {\em in parallel}.}
\RTHedit{Our} toolbox also simplifies setting up and solving multiphysics inversions in which data from different forward problems are jointly inverted.
The software contains three main components. The first is a mesh module that provides a set of routines for discretizing commonly used differential operators on a regular or stretched meshes.
The mesh module allows the user to experiment with different mesh types and quickly implement different forward problems.
It is accompanied by interfaces to highly efficient direct and iterative linear solvers for the solution of the forward problem that yield Fortran-like speed.
The second key component of \jInv{} is an inversion module for fitting the data given a particular forward problem.
It includes commonly used options for misfit and regularization functions as well state-of-the-art methods for constrained optimization.
The third component in this package is a module that enables parallel solution of the forward problems. It allows for different parallelization strategies and supports various computational architectures ranging from a multicore laptop, large shared memory servers, to heterogeneous platforms like cloud computing engines.

\RTHedit{Our code is implemented and documented in Julia and our core package \jInv{} can be obtained under the permissive MIT license at:
\begin{center}
	\texttt{https://github.com/JuliaInv/jInv.jl}
\end{center}
Our code repository includes a growing collection of tutorials, which highlight and explain key concepts in \jInv, and inversion examples, which demonstrate how \jInv{} is used to solve the parameter estimation problems in DC resistivity, FWI, and travel time tomography that are presented in this paper. The tutorials and examples consist of code, visualization, and documentation and can be viewed online. However, they can also be modified and executed on a local computer or online using services such as \texttt{JuliaBox}\footnote{http://www.juliabox.org}.
The \jInv{} code contains unit-tests and is automatically tested on different operating systems and architectures. The parallel code is tested on multicore laptops,  workstations, and cloud computing platform.
}

\RTHedit{There are a few alternative packages for PDE parameter estimation. Efficient and scalable algorithms can be created using  the more general scientific computing packages PETsc~\cite{balay2014petsc} and Trilinos~\cite{heroux2005overview}, and, in a more convenient way, through the Trilinos extension for PDE constrained optimization Sundance~\cite{long2012sundance}. However, those packages are written in C/C++, require sophisticated installation and are less suitable for rapid prototyping.
An alternative is the Python-based package SimPEG~\cite{CockettEtAl2015}, which is easier to install and simplifies designing of and experimenting with parameter estimation algorithms. However, SimPEG does not provide methods for parallelization and is, thus, most useful for solving small-scale problems. The FeniCS project~\cite{alnaes2015fenics} provides a variety of tools for automatic solution of a wide range of PDEs. Finite element discretization of linear and nonlinear PDEs and their adjoints that can also be  built using Python. FeniCS features parallel solvers for the PDE forward problems and there is an extension for PDE-constrained optimization~\cite{funke2013framework}. However, the implementations of the optimization algorithms described in~\cite{funke2013framework} are not designed for distributed execution on a cluster.
Our goal is to \emph{both} allow researchers to easily explore new ideas and develop new algorithms \emph{and} enable  practitioners to efficiently solve realistic large-scale parameter estimation problems in a common framework. These two objectives motivate our choice of the dynamic high-level programming language Julia and guide the development of \jInv. A few of the recent works that use our package and further demonstrate its use beyond this paper include \cite{belliveau2016achieving,haber2016solving,haber2016obtaining,mcmillan20163d,mcmillan2016multiple,treister2016full}}.

The remainder of the paper is organized as follows. In Sec.~\ref{sec2} we discuss the type of problems that can be
solved using our package. In Sec.~\ref{sec:inv} we  give a brief overview on computational methods for solving inverse problems with emphasis on efficient computations involving sensitivity matrices.
Sec.~\ref{sec:parallel} gives a detailed presentation of the parallel schemes in \jInv, which are the main contribution of this paper. We discuss design choices needed to obtain efficient parallelization in PDE parameter estimation problems and develop and compare two different approaches for parallelization.
In Sec.~\ref{sec:package_structure} we give a brief overview of the package structure.
The potential of \jInv\  is demonstrated using numerical experiments in Sec.~\ref{sec:applications}.
 Finally, in Sec.~\ref{sec6} we conclude the paper and discuss future direction of research.

\section{Supported forward problems}
\label{sec2}

In this section, we introduce the mathematical framework and the type of forward problems
that are supported in \jInv. These problems are all based on \RTHedit{PDEs} that depend on given parameters
(later to be assumed unknown). In the forward problem we assume that
all \RTHedit{the} parameters are known and solve the \RTHedit{discretized PDEs in order to simulate the data}. We conclude the section with an extensive discussion on direct and iterative methods provided by \jInv\ for solving discretized linear PDEs.

To be more specific,
let $\Omega \subset \R^d$ denote the computational domain where
the space dimension is either $d=2$ or $d=3$. In this paper, we consider the discrete inverse problem of estimating a model function $m : \Omega \to\R$ from noisy measurements
\begin{equation}\label{eq:data}
	\bfd_{ijk} = (p_{i},u_{jk}) + \epsilon_{ijk}, \quad \text{ for } \quad i=1,\ldots,n_p,\ j=1,\ldots,n_{\omega},\ k=1,\ldots,n_q,
\end{equation}
where the discrete measurements, $\bfd_{ijk}$, are defined by the $L_2$ inner product of the $i$th receiver function $p_i$, the field $u_{jk}$ and the measurement noise $\epsilon_{ijk}$. \RTHedit{The receiver functions describe the measurement devices used to take measurements of the field, e.g.,  Dirac delta functions that measure the field at some points in the domain or its boundary.}
The indices $j$ and $k$ indicate the dependence of the fields in frequency (or potentially time) and source location.
We assume that for a given model $m$, the fields satisfy
\begin{equation}
\label{problemsNonLinear}
{\cal F}(m,\omega_{j},u_{jk}) = q_{k},
\end{equation}
where we assume in this paper that ${\cal F}$ involves Partial Differential Equations (PDEs) for $u_{jk}$ accompanied with some boundary conditions. The PDEs may depend on the known frequency,
$\omega_k$,  and the source, $q_{k}$.
In this paper, we consider both the non-linear and the linear case. In the latter case the forward model simplifies to
\begin{equation}
\label{problems}
{\cal F}(m,\omega_{j},u_{jk}) = {\cal A}(m,\omega_{k}) u_{jk}  =  q_{j},
\end{equation}
\RTHedit{where $\CA(\cdot,\cdot)$ represents the PDE operator.}

This setting, using the linear problem type~\eqref{problems}, covers geophysical applications such as DC resistivity~\cite{deymor}, electromagnetic imaging~\cite{haberBook2014}, gravity and magnetics inversion, FWI~\cite{pratt1999,EpanomeritakisAkcelikGhattasBielak2008,krebs09ffw,VirieuxOperto2009,WarnerEtAt2013} and single phase flow in porous media, among other problems. The travel time tomography problem~\cite{sei1994gradient,leung2006adjoint,taillandier2009first,li2013first} is an example of a non-linear forward problem and based on the eikonal equation.
Similar applications in medical imaging are, e.g., in Diffuse Optical Tomography~~\cite{Arridge1999, ArridgeSchotland2009,SaibabaEtAl2015} or Electrical Impedance Tomography~\cite{CheneyEtAl1999}.
In the forward problem we compute $\bfd$ given $m$ and in the inverse problem we attempt to recover $m$
given the data $\bfd$, sources $q_1,\ldots,q_{n_q}$, receivers $p_1,\ldots,p_{n_p}$ and frequencies $\omega_1,\ldots,\omega_{n_\omega}$ where relevant.

\bigskip

In this work we demonstrate the solution of the forward, and later the inverse problem of the following three model problems:

\begin{example}[Direct Current (DC) Resistivity]
	\label{ex:divsiggrad}
DC resistivity is of major interest in applied geophysics~\cite{mcgillvery}.
The model function, $m$, parametrizes the conductivity $\sigma$ of the subsurface.
For this problem the governing equation is
\begin{equation}
\label{dc}
\div (\sigma(m(x)) \nabla u_{j}(x)) = q_{j}(x), \quad  x \in \Omega, \quad \nabla u_{j}(x) \cdot {\vec n} = 0,
\quad  x \in \partial \Omega.
\end{equation}
Here, $u_{j} : \Omega \to \R$ is the potential field that evolves from the $j$th source, $q_{j}$, which is
a dipole placed on the earth's surface.
A common parametrization is $\sigma(m) = \exp(m)$, which ensures positivity of the conductivity and can deal
with large dynamical ranges of conductivities.
Data is typically collected on the surface as well and is a difference between the potential field
$u_{j}$ at two points. The goal of the inverse problem is to recover $m$ from surface measurements of the potential difference field $\bfd \in \R^{n_p \times n_q}$.
\end{example}

\bigskip

A second example that we use is the seismic full waveform inversion.
\begin{example}[Full Waveform Inversion (FWI)]
	\label{ex:fwi}
Full waveform inversion is used for the evaluation of the velocity of wave propagation in the earth.
For this problem (in frequency domain), the governing equation is the Helmholtz equation
\begin{equation}
\label{fwi}
\begin{split}
	\div (\rho(x)^{-1} \nabla\, u_{jk}(x)) + \omega_{k}^{2}\, (1+\imath \gamma(x))\,  m(x)\,  u_{jk}(x)  = q_{j}(x), \quad  x \in \Omega,\\
	\quad \nabla u_{jk}(x) \cdot {\vec n} = 0, \quad  x \in \partial \Omega.
\end{split}
\end{equation}
Here, $\rho: \Omega \to \R$ is the density, and $u_{jk}:\Omega \to \mathbb{C}$ is the pressure field that evolves from the $j$th source $q_{j}$, and the $k$th frequency, $\omega_{k}$. Here, the model $m$ is the compressibility of the subsurface. The attenuation $\gamma : \Omega \to \R$ is assumed to be known and is also used to suppress artificial reflections from the boundary of the computational domain using an absorbing layer~\cite{liao1996multifrequency}. Data measurements are typically collected on the earth's surface and in boreholes and
the goal of the inverse problem is to recover $m$ given the measurements
of $\bfd$ on the surface.
\end{example}

\bigskip

As a third example, we consider travel time tomography that involves the eikonal equation and thus leads to a non-linear PDE constraint in~\eqref{problemsNonLinear}.
\begin{example}[Travel Time Tomography]
	\label{ex:traveltime}
Similarly to FWI, travel time tomography is also used to evaluate of the velocity of wave propagation in the earth. It is a simplified model of the wave propagation that considers only the first arrival time of the seismic waves. In comparison, FWI also includes reflections of the waves from different ground layers. In this paper, we consider the eikonal equation as a forward problem
\begin{equation}\label{eikonal}
|\nabla u_j(x)|^2 = m(x), \quad x \in \Omega, \quad u_j(x_j) = 0.
\end{equation}
Here, $|\cdot|$ is the Euclidean norm in $\R^d$, $u_{j} : \Omega \to \R$ is the arrival time of the first wave that evolves from the $j$th source $q_{j}$ that is located at $x_j \in \Omega$, and the model $m$ is the same compressibility parameter as in \eqref{fwi}, also known as the squared slowness. The first arrival time of the wave is typically extracted from the waveform data used in FWI. The goal of the inversion is to recover a rough estimate of $m$ given the measurements of $\bfd \in \R^{n_p \times n_s}$. Compared to FWI, travel time tomography is much easier computationally and in some cases it may also be more robust in terms of the obtained recovery.
\end{example}

\bigskip

\jInv\ provides numerical methods for evaluating the forward models on stretched tensor meshes. In the remainder of the paper, $\bfm \in \R^{N_m}$ denotes a discretization of the model function $m$ on a $d$-dimensional tensor mesh with $N_m$ cells. \RTHedit{For simplicity, let} us consider the computation of $\bfd_{ijk}$ for \RTHedit{some fixed indices} $i$, $j$, and $k$.
The field $u_{jk}$ and its corresponding source $q_j$ are discretized on a tensor mesh with $N_{jk}$ cells and are denoted by $\bfu_{jk}$ and $\bfq_j$, respectively. \RTHedit{Each discrete field $\bfu_{jk}$} satisfies a discrete version of~\eqref{problemsNonLinear}, i.e.,
\begin{equation}
	\label{eq:probNLdiscrete}
	\bfF_{jk} (\bfm, \bfu_{jk};\omega_{k}) = \bfq_{j},
\end{equation}
where $\bfF_{jk}: \R^{N_m} \times \R^{N_{jk}} \to \R^{N_{jk}}$ is assumed to be continuously differentiable in its first two arguments.
The case when $N_m \neq N_{jk}$, also known as \emph{mesh decoupling}, allows the user to employ, e.g., a coarse discretization of the fields and a fine discretization of the model; see~\cite{HaberSchwarzbach2014} for more details and Sec.~\ref{sub:3d_joint_travel_time_and_dc_resistivity_survey} for an example.
Assuming that the discrete problem has a unique solution for the models of interest, we denote $\bfu_{jk}(\bfm)$.

\jInv\  provides several options for solving~\eqref{eq:probNLdiscrete} given a model $\bfm$. For example, for the  DC resistivity problem,~\eqref{eq:probNLdiscrete} is linear in $\bfu$ and reads
\begin{equation}\label{eq:probLinDiscrete}
    \bfF_{j} (\bfm, \bfu_{j}(\bfm)) = \bfA_j(\bfm) \bfu_{j}(\bfm) = \bfq_j \quad \Rightarrow \quad \bfu_{j}(\bfm) = \bfA_j(\bfm)^{-1} \bfq_j,
\end{equation}
where $\bfA_{j}(\bfm) \in \R^{N_j \times N_j}$ is a sparse and symmetric positive definite matrix obtained from a mimetic finite volume discretization of~\eqref{dc}. The choice of an effective solver for this linear system of equations depends on the problem size and computational resources.

\subsection{Solvers for linear forward problems}\label{sec:LinearSolvers}
The solver for linear systems is a modular component in \jInv, where we provide several options using a single interface. This interface can be used with a shared memory version of the sparse direct solver MUMPS~\cite{MUMPS}, compiled with metis~\cite{METIS} as re-ordering scheme, via the Julia package \href{https://github.com/JuliaSparse/MUMPS.jl}{\texttt{MUMPS.jl}}. Another direct solver that is interfaced in our package is \cite{schenk2004solving}, via the Julia package \href{https://github.com/JuliaSparse/Pardiso.jl}{\texttt{Pardiso.jl}}. These options are most suitable for 2D or relatively small 3D problems. The same interface is also used for iterative Krylov solvers, such as (Preconditioned) Conjugate Gradient (PCG)~\cite{hest} and Biconjugate gradient stabilized (BiCGSTAB)~\cite{van1992bi}.
We support both the standard and block versions of these methods~\cite{OLeary1980,el2003block}, via the package \href{https://github.com/lruthotto/KrylovMethods.jl}{\texttt{KrylovMethods.jl}}.
The block versions are especially important for exploiting the shared-memory parallelism of the machine, via parallelized sparse matrix times dense matrix product, which is implemented in Fortran using OpenMP in the package \href{https://github.com/JuliaInv/ParSpMatVec.jl}{\texttt{ParSpMatVec.jl}}. In addition, we use level-3 BLAS routines available in Julia for parallelizing the dense matrix products.

When using iterative methods, the choice of an effective preconditioner is key since many linear solves with relatively high accuracy are required. In \jInv, we use Symmetric Successive Over-relaxation (SSOR)~\cite{Saad2003} by default as this is a generic choice for many problems. Alternatively, we provide in \jInv\ some more sophisticated multigrid preconditioners \RTHedit{given in the package \href{https://github.com/JuliaInv/Multigrid.jl.git}{\texttt{Multigrid.jl}}. These} can be incorporated as block preconditioners with the block Krylov methods mentioned above for multiple right-hand-sides (in addition to standard preconditioning). Similarly to before, this is important for exploiting shared memory parallelism in the multigrid relaxations (sparse matrix-matrix products) and direct solution of the coarsest grid problem.

For the DC resistivity forward problem we use an implementation of the Smoothed Aggregation algebraic multigrid algorithm described in~\cite{treister2015non}, which can solve the linear system~\eqref{dc} for a highly heterogenous conductivity $\sigma$. Using PCG with this preconditioner, we achieve mesh-independent convergence for this problem. Since the preconditioner is algebraic, in principle this combination can solve the problem for any mesh. For the Helmholtz linear system in~\eqref{fwi}, we have implemented a variant of the shifted Laplacian preconditioner~\cite{erlangga2006novel,oosterlee2010shifted,airaksinen2007algebraic} using a geometric multigrid framework on a regular grid. This problem is considered harder than the previous one, as it is indefinite. We note that the user can easily extend the linear solvers module by interfacing other solvers.

\section{Computational methods for inversion} 
\label{sec:inv}

This section briefly discusses the key components that we use for solving the PDE parameter estimation problems introduced in the previous section. Particular emphasis is given to computing sensitivities,
but we also review common choices for misfit functionals, regularization, and numerical optimization that are implemented in \jInv. The interested reader is referred to~\cite{Haber} for a detailed discussion.

To estimate the model $\bfm$ given the data $\bfd$, we form and solve
the reduced optimization problem\footnote{\RTHedit{In this paper we refer to the reduced optimization problem where $\bfu_{jk}$ is defined as a function of $\bfm$, $\bfu_{jk}(\bfm)$. The original PDE-constrained optimization problem includes a minimization over $\bfm$ and all the fields $\bfu_{jk}$, with \eqref{eq:probNLdiscrete} as constraints. }}
\begin{equation}
	\label{genOpt}
		\min_{\bfm}  \hf \sum_{ijk} \phi((\bfp_{i}, \bfu_{jk}(\bfm)),\bfd_{ijk} ,\bfw_{ijk}) + \alpha R(\bfm)\quad
		\text{ subject to }\quad  \bfm_{\rm L} \leq \bfm \leq \bfm_{\rm H},
\end{equation}
where $\bfu_{jk}(\bfm)$ are the fields that satisfy \RTHedit{\eqref{eq:probNLdiscrete} or \eqref{eq:probLinDiscrete}}. Here, $\bfw_{ijk} \in \R^+$ are weights that corresponds to the inverse standard deviation of each datum, $\phi$ is a misfit function, $R(\bfm)$ is a regularization function, and $\bfm_{\rm L} \in \R^{N_m}$ and $\bfm_{\rm H} \in \R^{N_m}$ are lower and upper bounds on the model, respectively. For the misfit functions $\phi$ we consider either a weighted $\ell_2$-norm, or a smooth approximation of the $\ell_1$-norm. In addition, we provide a smoothing and total variation regularizers~\cite{vogelBook}. The regularization parameter $\alpha\geq0$ balances between minimizing the misfit functional and the regularizer and must be chosen by the user.

Since the problems that are considered here are nonlinear (in the model),
a key ingredient of efficient inversion algorithms is computing the sensitivity matrix
(or Jacobian), $\bfJ_{jk}(\bfm) \in \R^{N_{jk} \times N_m}$, that describes the change of the
field with respect to small perturbations of the model. To be precise, for any small perturbation $\delta\bfm \in \R^{N_m}$ the sensitivity matrix $\bfJ_{jk}$ satisfies
\begin{equation}
\label{eq:TaylorApprox}
	\bfu_{jk} (\bfm + \delta\bfm) \approx \bfu_{jk} (\bfm) + \bfJ_{jk}(\bfm) \delta\bfm + {\cal O}(\|\delta\bfm\|^{2}).
\end{equation}
The sensitivity matrix is derived by differentiating both sides of the discretized forward problem~\eqref{eq:probNLdiscrete} with respect to $\bfm$, which yields
\begin{equation*}
0 = \nabla_\bfm \left( \bfF_{jk} (\bfm, \bfu_{jk}(\bfm)) \right) = \nabla_\bfm \bfF_{jk} (\bfm, \bfu_{jk}) + \nabla_{\bfu_{jk}}  \bfF_{jk} (\bfm, \bfu_{jk})  \bfJ_{jk}(\bfm)\\
\end{equation*}
and assuming that $ \nabla_{\bfu_{jk}}  \bfF_{jk} (\bfm, \bfu_{jk})$ is invertible we obtain
\begin{eqnarray}\label{eq:sens}	
\bfJ_{jk} (\bfm) = - \left( \nabla_{\bfu_{jk}}  \bfF_{jk} (\bfm, \bfu_{jk})\right)^{-1} \ \nabla_\bfm \bfF_{jk} (\bfm, \bfu_{jk}).
\end{eqnarray}
In the special case of~\eqref{eq:probLinDiscrete}, the sensitivity is given by
\begin{equation*}
	\bfJ_{jk}(\bfm) = - \bfA_j(\bfm)^{-1}  \ \nabla_\bfm \left( \bfA_{j} (\bfm) \bfu_{jk} \right).
\end{equation*}

It is important to note that while the sensitivity matrix is generally dense, its product (or the product of its adjoint) with a vector can be computed by solving a (linearized) forward problem (or its adjoint), respectively.
By default, sensitivity matrices are not build explicitly in \jInv{} to save computation time and memory. Instead, methods for computing matrix-vector products $\bfJ_{jk}(\bfm) \bfv$ and $\bfJ_{jk}(\bfm)^\top \bfw$ are provided. Depending on the linear solver being used, temporary results such as fields, factorizations or preconditioners are stored and re-used to speed up computations involving the sensitivity matrix. Computing the sensitivities for nonlinear PDE forward problems is along the same lines. For the travel time tomography problem see, e.g.,~\cite{TreisterHaber2016}.

The bound-constrained optimization problem~\eqref{genOpt} is solved using the projected Gauss-Newton method also described in~\cite{haberBook2014}. In each iteration, the approximation~\eqref{eq:TaylorApprox} is placed in~\eqref{genOpt} for $\bfu_{jk}$, resulting in a convex quadratic objective for $\delta\bfm$. On the active set, the update $\delta\bfm$ is computed by projected steepest descent and on the inactive set by computing a Gauss-Newton update using a projected Preconditioned Conjugate Gradient (PCG) method; see~\cite{nw}.
Note that each iteration of the projected PCG method requires one matrix vector multiplication with the sensitivity matrix and one with its transpose for each source and frequency, which sums up to $2 \cdot n_q \cdot n_{\omega}$ PDE solves.
Note that the model and thus the PDE coefficients do not change within each Gauss-Newton iteration and thus temporary results from the PDE solves can be re-used to accelerate the inner solve; see also remark below.
A projected Armijo backtracking is used as a line search criteria.

\begin{remark} In addition to the current model, $\bfm$, computations with the sensitivity matrix also require the corresponding fields $\bfu_{jk}$ as can be seen in~\eqref{eq:sens}.
Storing the fields in memory can reduce runtime considerably, however, it can be prohibitively memory-consuming in large-scale inversions with a large number of sources and frequencies. In our code, we keep the fields in memory with a low precision by default (using 32-bit single precision, for example). Furthermore, we have the option to write the fields to disk and use them in small batches of sources. Because the results of applying the sensitivity involve a sum over all frequencies and sources, having the fields in batches is manageable in memory, and we are able to treat multiple sources and frequencies. The fields are compressed using the HDF5 format \cite{hdf5}, which is also used in MATLAB and is available in Julia using the package {\tt MAT.jl}.
\end{remark}

\section{Parallel PDE parameter estimation}
\label{sec:parallel}

\jInv\  provides different ways to use parallel computing to accelerate solving the optimization problem~\eqref{genOpt}. In practice, the choice of a successful parallelization strategy depends on many factors related to the experimental setup and the available computational resources. Thus, a key goal of \jInv\  is to provide flexibility to the user to pick or mix different strategies. In this section we discuss a number of options to parallelize the problem, their advantages and limitations as well as possible use cases.

Typically, the main computational bottleneck in PDE parameter estimation problems is  solving the forward problems which requires numerical solution to many PDEs. Thus, a straightforward way to reduce computation times is to use a state-of-the-art parallel PDE solvers. For example, if the governing PDE is linear, parallelized direct methods can be used for factorizing the discretized PDEs~\eqref{eq:probLinDiscrete}. As described in Sec.~\ref{sec:LinearSolvers}, we provide a wrapper to the highly efficient sparse solvers MUMPS~\cite{MUMPS} and Pardiso~\cite{schenk2004solving} with shared memory parallelization. This option is attractive, e.g., when the same mesh and frequency is used to solve all forward problems and the size of the linear system is small enough given the available memory or disk space. 
A key benefit is that, once the factorization is available, matrix-vector products with the sensitivity matrix can be computed relatively cheaply.
Direct solvers can be prohibitively expensive, e.g., when using a very fine discretization of the forward problems or when using mesh decoupling. In this case, iterative methods are often preferred, and to exploit shared memory parallelism we provide a library for parallelized sparse matrix vector products; see also Sec.~\ref{sec:LinearSolvers}. However, there are cases were the PDE solver cannot easily be parallelized, e.g., in travel time tomography; see Example~\ref{ex:traveltime}.

In addition to using parallel PDE solvers, we can exploit the fact that the terms in the misfit function in~\eqref{genOpt} can be divided into small batches that can be computed independently.
In fact, most geophysical experiments are involved with multiple and potentially many sources and frequencies, often, a massive
number of them.
Thus, parallelization is possible by simulating PDEs associated with different sources, receivers, or frequencies in parallel.
 While this may appear to be trivial in principle, it is more complex in practice and requires some careful design of the inversion software as well as the discretization of the forward problem.
While in principle all PDE simulations can be evaluated independently, we might still want to group some of the problems together to gain efficiency. For example, when using a direct linear solver, the main cost is factorizing $\mathrm{A}(\bfm,\omega_k)$ for each frequency. Thus, it might be beneficial to group all source, receiver, frequency combinations by their frequencies and parallelize over different frequencies.
More sophisticated strategies require the design of different meshes for every source-receiver combination; see~\cite{HaberSchwarzbach2014} for details. \jInv\  supports all these different designs and allows the user
to change the parallelization strategy based on the problem at hand.

In the remainder of this section, we describe two options for asynchronous parallel solution of the forward problems. First, we describe an on-the-fly parallelization with dynamic scheduling that aims at minimizing latencies. Second, we describe a distributed memory version that reduces communication costs. The latter case uses a static assignment of problems to workers, i.e., the forward problems are distributed among the workers a priori and the assignment is then being kept fixed. Finally, we present our parallel optimization method that supports both dynamic and static scheduling. In the following, we assume that the forward problems are divided into a $n_b$ smaller batches and that, without loss of generality, $n_b$ is greater or equal than the number of available workers denoted by $n_W$.

\subsection{Dynamic scheduling for asynchronous computation of the forward problems}

Let us begin with the most common approach for solving the forward problems in~\eqref{genOpt} in parallel. This approach is similar techniques used \RTHedit{for solving multi-stage stochastic programming problems, e.g., in two-stage linear programming \cite{LinderothWright2003} and in power grid optimization using Julia~\cite{HuchetteEtAl2014}}.

In order to solve the forward problems asynchronously, we send the model and one batch of forward problems to each available workers. Then each worker solves its respective forward problem locally and communicates the result to the main process. The main process then sends the next available batch to this worker until all batches have been assigned.
Thereby the number of idle processors is kept to a minimum.
The dynamic scheduling approach for computing the misfit reads is summarized in Alg.~\ref{alg:dynamic}.

 \begin{algorithm}[t]
\caption{Dynamic Scheduling for Computing Forward Problems}
	\label{alg:dynamic}
	   \begin{algorithmic}
		\STATE create list of batches $J=[1,2,\ldots,n_b]$
		\FOR{$k=1,\ldots,n_W$ (asynchronously and in parallel)}
			\WHILE{$J$ is not empty}
				\STATE draw first element $j$ from $J$ and remove it from $J$
			 \STATE send data for $j$th batch of forward problems to worker $k$
			 \STATE worker solves PDEs and computes data
			 \STATE when done: main process fetches simulated data and fields
			\ENDWHILE
		\ENDFOR
	 \end{algorithmic}
\end{algorithm}

There are some obvious advantages to this approach.
\RTHedit{First, it is simple to implement. Second, it can be efficient if sending forward problems to workers and retrieving their results does not involve a lot of communication.} However, in contrast to many other distributed optimization problems, computing the forward problems \RTHedit{in our case} often requires a non negligible amount of data such as sources, receivers, and meshes.
When using dynamic scheduling we keep sending forward problems to a respective worker, which causes significant amount of communication. The amount of communication can be decreased by re-computation
of different quantities whenever they are called, however, this may slow down the process
significantly and make the approach non-competitive.

\subsection{Static scheduling for asynchronous computation of the forward problems}
\label{sub:static}
An alternative approach, aimed at reducing  communication costs, is to distribute the forward problems to the workers once and then communicate the current model only.
This is done in two stages. First, the forward problem data is distributed among all available workers and each worker prepares the problems, e.g., by building PDE operators; see Alg.~\ref{alg:static1}. This step is performed asynchronously and a dynamic strategy similar to Alg.~\ref{alg:dynamic} is used to map problems to workers.
After this setup phase, the misfit can be computed asynchronously and in parallel; see Alg.~\ref{alg:static2}. Computing the misfit requires only the communication of the current model from the main processor to the workers. By default, the computed data and fields are not sent back to the main process, but a remote reference is provided.
\begin{algorithm}[t]
\caption{Static parallelization - Step 1: distributing forward problems}
\label{alg:static1}
   \begin{algorithmic}
     	\STATE create list of batches $J=[1,2,\ldots,n_b]$
	  	\FOR{$k=1,\ldots,n_W$ (asynchronously and in parallel)}
			\WHILE{$J$ is not empty}
				\STATE draw and remove first element $j$ from $J$. Set $\bfM_j=k$
				\STATE send data for $j$th batch of forward problems to worker $k$
			 	\STATE worker \RTHedit{$k$} prepares differential operators
			 	\STATE when done: leave updated forward parameters on worker and continue
			\ENDWHILE
		\ENDFOR
	 \STATE Output the map $\bf M$ that maps forward problems to processors for further use.
	 \end{algorithmic}
\end{algorithm}
\begin{algorithm}[t]
\caption{Static parallelization - Step 2: computing the forward problems}
\label{alg:static2}
   \begin{algorithmic}
	 \STATE send current model, $\bfm$, to all processors (asynchronously)
		\FOR{$k=1,\ldots,n_W$ (asynchronously and in parallel)}
			\FOR{$j=1,\ldots,n_b$ (loop over all batches)}
				\IF{$\bfM_j$ is equal to $k$ ($j$th batch is stored on worker $k$)}
					\STATE worker $k$ solves PDEs and computes data
			 		\STATE when done: return remote reference to simulated data and fields
				\ENDIF
			\ENDFOR
		\ENDFOR
	 \end{algorithmic}
\end{algorithm}

The static assignment is attractive as it reduces the amount of communication by avoiding sending forward problems and data to different workers when the model is updated and new data needs to be simulated.
However, latency times might be greater as compared to dynamic scheduling if the initial assignment does not properly balance the computation between the workers.

The choice of static versus dynamic implementation is highly problem and system dependent. In some cases, where communication is very fast or when the memory is shared, the dynamic allocation can perform better than the static one. On the other hand, we have found that for most problems of interest, where the amount of data to be transferred is non-trivial and communication between main process and workers is slow, the static implementation
can yield faster results. \RTHedit{\jInv\  supports both modes of computation and allows the user to choose the approach that best fits the problem at hand.}

\subsection{Parallel Gauss-Newton} 
\label{sub:parallel_optimization_methods}
With a parallel method for solving the forward problems at hand we can evaluate the objective function and its gradient in parallel. Thus, deriving a parallel first-order optimization method is straightforward.

In our experience, however, first-order methods converge slowly and thus lead to a considerably greater number of function evaluations (and thus PDE solves). We therefore favor a projected Gauss-Newton method since it typically needs fewer iterations, and enables us to save the computation time of the sensitivity by reusing the temporary results (factorizations and preconditioners) obtained for evaluating the misfit. Storing the fields and these temporary results from the PDE simulations can dramatically reduce the cost of the sensitivity computations, which dominate the cost of the Gauss-Newton algorithm. Similar approaches have been used, e.g., in in~\cite{EpanomeritakisAkcelikGhattasBielak2008,HaberSchwarzbach2014}.

We compute the Gauss-Newton search direction using a matrix-free projected PCG. The individual terms of \RTHedit{the} gradient and the Hessian are computed in parallel across all workers according to the specified scheduling. Each worker computes its terms of the Hessian matrix vector product and communicates only the aggregated result. In order to avoid unnecessary computations and extensive communication between the workers, the temporary results obtained in the misfit calculation for solving a forward problem $i$ need to be readily available on the specific remote worker that is scheduled to solve the problem $i$ in the sensitivity calculation. This rule is naturally preserved when using the static scheduling in Algorithms \ref{alg:static1}-\ref{alg:static2}, since the assignment of problems to workers is fixed. However, the rule is not preserved in the case of dynamic scheduling, as the assignment of the forward problems to workers can change between one sensitivity calculation to the next. In this case, in our implementation we keep the rule by fixing the assignment of forward problems to workers within a Gauss-Newton iteration after evaluating the misfit function. Thus, intermediate results are stored at the respective workers to accelerate Hessian matrix vector products. The assignment is re-computed in next misfit calculation in the line search procedure after obtaining the Gauss-Newton direction.

\section{Package structure} 
\label{sec:package_structure}

This section outlines the main structure of \jInv\  and lists additional resources. Detailed up-to-date instructions about installing and using \jInv\  can be found on its Github web-page. 

Following the common practice of Julia programming, the \jInv\  code is represented as a module with \RTHedit{six} submodules. Roughly speaking, a module in Julia is a collection of code that defines its own namespace, which helps prevent conflicts at run time. A submodule is a module inside a module and can be loaded individually. In \jInv\  we think about submodules as individual building blocks that can be combined as needed to solve real-world problems. To enforce a clear structure we organize each module in a separate folder, and try to minimize the dependency between the modules. \jInv\  consists of the following submodules:
\begin{itemize}
	\item \texttt{Mesh} - regular and stretched mesh as well as discretization of commonly used differential and integral operators on these meshes. This abstract interface allows the user to easily test different discretization strategies and even use different meshes for each batch of forward problems.
	\item \texttt{LinearSolvers} - provides an interface for solving linear PDEs such as~\eqref{problems}, as described in Sec.~\ref{sec:LinearSolvers}.
	\item \texttt{InverseSolve} - provides type used to represent the misfit and other inversion parameters. Also provides  methods for solving the inverse problems such as misfit functions, regularization, model functions, and optimization; see~\eqref{genOpt}.
	\item \texttt{ForwardShare} - methods for parallel evaluation of forward problems. User-defined forward problems that respect the structure defined here, can automatically benefit from our parallelization options.
	\item \RTHedit{\texttt{Vis} - functions for visualization and plotting.}
	\item \texttt{Utils} - some helper functions and testing routines.
\end{itemize}
All our modules declare their dependencies, provide documentation, and automatic testing. \RTHedit{Our code repository contains interactive tutorials and inversion examples.} The implementation of the three examples of inverse problems in Section \ref{sec:applications} are given in the packages \href{https://github.com/JuliaInv/DivSigGrad.jl}{\texttt{DivSigGrad.jl}}, \href{https://github.com/JuliaInv/FWI.jl}{\texttt{FWI.jl}}, and \href{https://github.com/JuliaInv/EikonalInv.jl}{\texttt{EikonalInv.jl}}.

\section{Numerical experiments}
\label{sec:applications}

In this section we demonstrate the capabilities of \jInv{} for solving the example inverse problems presented in Section \ref{sec2}. In addition, we show some scalability tests for typical usage scenarios of the package.
An advantage of using Julia is the cross-platform support. While our code supports standalone computers or laptops as well, this section focusses on the parallel efficiency of the code. We use the current stable release version of Julia 0.4.6 that is accessed via Github and compiled on each system. Tests are performed using three different test platforms.
\begin{description}
	\item[\emph{\textmd{Laptop:}}] Standard laptop computer running Windows 10 64bit OS, with Intel core-i7 2.8 GHz CPU with 4 cores and 32 GB of RAM.
	\item[\emph{\textmd{Shared memory computer:}}] Larger workstation operating Ubuntu 14.04 with 2 $\times$ Intel Xeon E5-2670 v3 2.3 GHz CPUs using 12 cores each, and a total of 128 GB of RAM. Julia is installed and compiled using Intel Math Kernel Library (MKL).
	\item[\emph{\textmd{Cloud computing engine:}}] We use the Amazon EC2 cloud computing engine. A customized Amazon Machine Image (AMI) containing Julia and all modules required by \jInv\ was created based on the Ubuntu server volume. Instances are launched on demand and a machine file is used to connect Julia instances. For the experiments below, we use the \texttt{c4.large} machine with two virtual CPUs and working memory of 3.75GB.
\end{description}

\subsection{3D FWI survey} 
\begin{figure}
\begin{center}
	\newcommand{\image}[1]{\includegraphics[width=0.48\linewidth]{#1}}
  \subfigure[\footnotesize The SEG/EAGE velocity model.]{\image{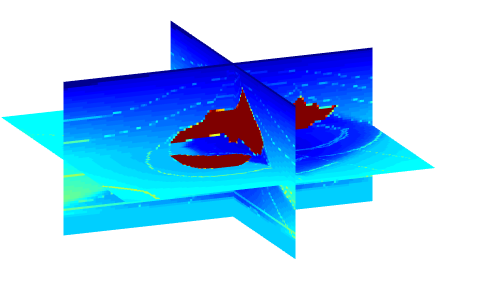}\label{fig:mtrue}}
  \subfigure[\footnotesize The starting linear model.]{\image{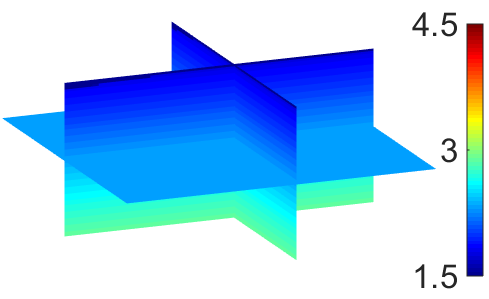}\label{fig:mref}}
    \subfigure[\footnotesize The FWI reconstruction.]{\image{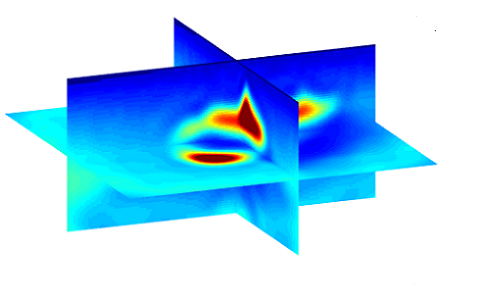}\label{fig:mfwi}}
\end{center}
\caption{\footnotesize Experimental setting for 3D full waveform inversion. Velocity units are in $km/sec$.}
\label{fig:SEG}
\end{figure}

We consider a synthetic Full Waveform Inversion (FWI) example using the 3D SEG/EAGE model~\cite{aminzadeh19973} as a ground truth model for the seismic wave velocity of the subsurface. The model, presented in Fig.~\ref{fig:mtrue}, contains a salt dome in which the velocity is significantly higher than in the background. The velocity ranges from around $1.5$ km/sec to around $4.8$ km/sec. The domain size is $13.5\,{\rm km}\times13.5\,{\rm km}\times4.2\,{\rm km}$. The domain of interest is divided into $145\times145\times70$ equally sized mesh cells of approximate size of $93\,{\rm m}\times93\,{\rm m}\times60\,{\rm m}$. To populate part of the absorbing boundary layer necessary for the forward problem in~\eqref{fwi} we add a 10-point padding for each boundary surface of the grid (except the free surface), resulting in a $165\times165\times80$ grid for the forward and inverse problems. We use a constant density $\rho=1$, and use the frequencies $f_i = \{0.5,0.75,1.25,1.75\}$ Hz, and set $\omega_i = 2\pi f_i$ in~\eqref{fwi}. We use 81 sources that are arranged on a $9 \times 9$ grid located at the center of the free surface, where the distance between each source is $1.488\,{\rm km}$. The waveform data is given at 16,641 receivers that are arranged on a $129 \times 129$ grid located at the center of the free surface, distanced $93\,{\rm m}$ from each other. The data $\bfd_i \in \C^{16,641 \times 81}$ is simulated on the same mesh for each frequency $f_i$. 

For the inversion, we apply a frequency continuation strategy, where we initially apply the Gauss-Newton algorithm for the lowest frequency 0.5 Hz, and gradually add higher and higher frequency data to the inversion. We use a single worker for the inversion, applying the direct method MUMPS to solve the forward problems~\eqref{fwi}. \RTHedit{We also observe similar results using multiple workers and BiCGSTAB with a shifted Laplacian multigrid preconditioner. }  In order to be able to simultaneously hold the factorization of the matrices in memory, we apply the continuation in batches of two consecutive frequencies at the time, and \RTHedit{cycle} through all the frequencies twice. Furthermore, when handling each frequency, the worker solves the Helmholtz systems for all the sources in batches of only 27 sources at the time, using the same factorization, to further reduce the memory footprint (the fields, which are necessary for the sensitivities, are saved to the disk). \RTHedit{The batches of the solutions are accumulated for calculating the sensitivities.} When inverting for each pair of frequencies, we apply 10 Gauss-Newton iterations with 9 projected PCG iterations in each. To enforce a smooth reconstruction, a diffusion regularizer is added. The regularization
parameter was chosen very small, i.e., $\alpha = 10^{-10}$, but the Hessian of the regularizer is used as a preconditioner, so the number of projected PCG iterations essentially takes the role of a regularization parameter. Lower and upper bounds of 1.5 and 4.5 $km/sec$ on the velocity model are chosen such that reconstructed model is physically plausible.

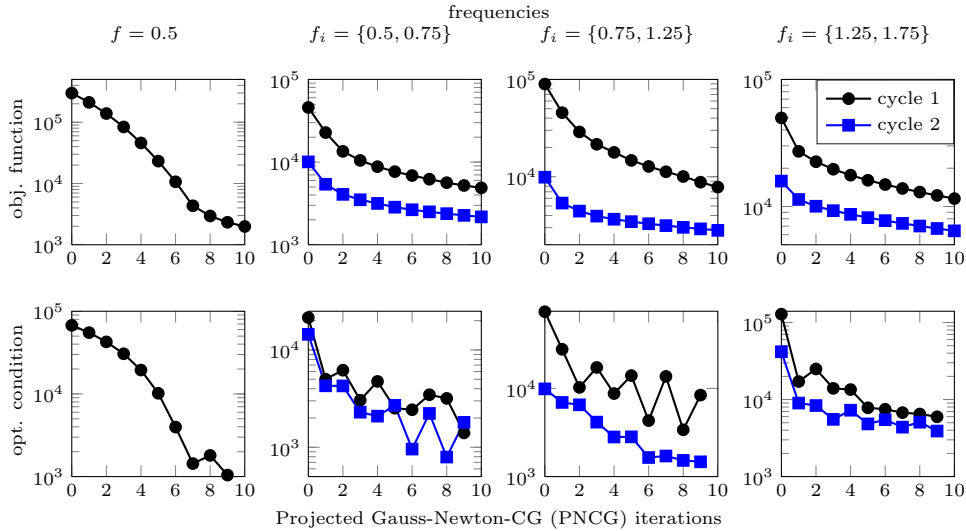
\begin{figure}[t]
	\begin{center}
		\newcommand{\rottext}[1]{\rotatebox{90}{\hbox to 30mm{\hss \scriptsize #1\hss}}}
		\iwidth=23mm
		\iheight=22mm
		\scriptsize
		
		\begin{tabular}{@{}c@{}c@{}c@{}c@{}c@{}}
			\scriptsize
			& \multicolumn{4}{c}{frequencies}\\
		&  $f=0.5$ & $f_i=\{0.5,0.75\}$ & $f_i=\{0.75,1.25\}$ & $f_i=\{1.25,1.75\}$\\
		\rottext{obj. function}   & \begin{tikzpicture}
\begin{axis}[%
width=\iwidth,
height=\iheight,
at={(0,0)},
scale only axis,
xmin=0,
xmax=10,
ymin=1000,
ymax=500000,
ymode=log,
yminorticks=true
]
\addplot [color=black,thick,solid,mark=*]
  table[row sep=crcr]{%
0   298850.615800      \\
1   210963.878255      \\
2   138688.325886      \\
3   83782.899993       \\
4   45928.402033       \\
5   23191.109798       \\
6   10668.655746       \\
7   4332.204868        \\
8   2949.397647        \\
9   2327.861947        \\
10  1987.465905        \\
};
\end{axis}
\end{tikzpicture}%
 & \begin{tikzpicture}
\begin{axis}[%
width=\iwidth,
height=\iheight,
at={(0,0)},
scale only axis,
xmin=0,
xmax=10,
ymin=1000,
ymax=100000,
ymode=log,
yminorticks=true
]
\addplot [color=black,thick,solid,mark=*]
  table[row sep=crcr]{%
0   45778.112087 \\
1   22753.855385 \\
2   13485.526511 \\
3   10474.566895 \\
4   8798.689799  \\
5   7667.561519  \\
6   6873.747733  \\
7   6212.788333  \\
8   5638.972622  \\
9   5210.297857  \\
10  4882.208446  \\
};
\addplot [color=blue,thick,solid,mark=square*]
  table[row sep=crcr]{%
0   10057.560628 \\
1   5393.505652  \\
2   4069.362457  \\
3   3501.667605  \\
4   3163.142738   \\
5   2848.000701   \\
6   2650.184051   \\
7   2501.065790   \\
8   2378.436243   \\
9   2267.555012   \\
10  2174.301367   \\
};
\end{axis}
\end{tikzpicture}%
 & \begin{tikzpicture}
\begin{axis}[%
width=\iwidth,
height=\iheight,
at={(0,0)},
scale only axis,
xmin=0,
xmax=10,
ymin=2000,
ymax=100000,
ymode=log,
yminorticks=true
]
\addplot [color=black,thick,solid,mark=*]
  table[row sep=crcr]{%
0   89599.047414 \\
1   45519.756601 \\
2   28872.314192 \\
3   21544.171180 \\
4   17796.699939  \\
5   14726.631650  \\
6   12756.444542  \\
7   11253.235032  \\
8   10048.422909  \\
9   8792.246330   \\
10  7830.627902   \\
};
\addplot [color=blue,thick,solid,mark=square*]
  table[row sep=crcr]{%
0  9891.345476 \\
1  5362.750957 \\
2  4434.932251 \\
3  3939.717893 \\
4  3659.143098  \\
5  3454.250669  \\
6  3287.898444  \\
7  3145.323036  \\
8  3021.942287  \\
9  2912.114248  \\
10 2813.432090  \\
};
\end{axis}
\end{tikzpicture}%
 &\begin{tikzpicture}
\begin{axis}[%
width=\iwidth,
height=\iheight,
at={(0,0)},
scale only axis,
xmin=0,
xmax=10,
ymin=5000,
ymax=100000,
ymode=log,
yminorticks=true,
legend style={at={(0.999,0.999)},anchor=north east,legend cell align=left,align=left,draw=white!15!black}
]
\addplot [color=black,thick,solid,mark=*]
  table[row sep=crcr]{%
0   49789.831179 \\
1   27159.880219 \\
2   22509.243727 \\
3   19654.897663 \\
4   17639.588978  \\
5   16085.657359  \\
6   14885.884521  \\
7   13866.729010  \\
8   12987.783667  \\
9   12218.133127  \\
10  11538.631542  \\
};
\addlegendentry{cycle 1}
 \addplot [color=blue,thick,solid,mark=square*]
   table[row sep=crcr]{%
 0  15850.02209  \\
 1  11356.50281  \\
 2  10048.57519  \\
 3  9273.290665  \\
 4  8677.329341   \\
 5  8172.258758   \\
 6  7743.058023   \\
 7  7363.855388   \\
 8  7022.314096   \\
 9  6714.528860   \\
 10 6434.027287    \\
 };
\addlegendentry{cycle 2}
\end{axis}
\end{tikzpicture}%
\\
		\rottext{opt. condition} & \begin{tikzpicture}
\begin{axis}[%
width=\iwidth,
height=\iheight,
at={(0,0)},
scale only axis,
xmin=0,
xmax=10,
ymin=1000,
ymax=100000,
ymode=log,
yminorticks=true
]
\addplot [color=black,thick,solid,mark=*]
  table[row sep=crcr]{%
0   67563.815124    \\
1   55142.607590    \\
2   42604.227493    \\
3   30590.768807    \\
4   19441.324402    \\
5   10179.288523    \\
6   3978.014843     \\
7   1436.167405     \\
8   1803.770091     \\
9   1043.054345     \\
};
\end{axis}
\end{tikzpicture}%
 & \begin{tikzpicture}
\begin{axis}[%
width=\iwidth,
height=\iheight,
at={(0,0)},
scale only axis,
xmin=0,
xmax=10,
ymin=500,
ymax=25000,
ymode=log,
yminorticks=true
]
\addplot [color=black,thick,solid,mark=*]
  table[row sep=crcr]{%
0   21529.834855   \\
1   5005.342720    \\
2   6198.219968    \\
3   3049.327572    \\
4   4751.316613    \\
5   2520.582608    \\
6   2429.769023    \\
7   3468.716545    \\
8   3180.466675    \\
9   1405.287228    \\
};
\addplot [color=blue,thick,solid,mark=square*]
  table[row sep=crcr]{%
0   14478.064226   \\
1   4281.329344    \\
2   4272.281681    \\
3   2294.633671    \\
4   2089.508796    \\
5   2695.739844    \\
6   958.832075     \\
7   2226.811741    \\
8   794.306403     \\
9   1804.359040    \\
};
\end{axis}
\end{tikzpicture}%
 & \begin{tikzpicture}
\begin{axis}[%
width=\iwidth,
height=\iheight,
at={(0,0)},
scale only axis,
xmin=0,
xmax=10,
ymin=1000,
ymax=75000,
ymode=log,
yminorticks=true
]
\addplot [color=black,thick,solid,mark=*]
  table[row sep=crcr]{%
0   73977.988191 \\
1   27752.610146 \\
2   10255.264961 \\
3   17280.225091 \\
4   8740.890867  \\
5   13974.225919 \\
6   4315.569938  \\
7   13675.055288 \\
8   3394.299331  \\
9   8395.448210  \\
};
\addplot [color=blue,thick,solid,mark=square*]
  table[row sep=crcr]{%
0   9872.681456  \\
1   6933.338889  \\
2   6515.118697  \\
3   4145.541144  \\
4   2811.547990  \\
5   2821.893371  \\
6   1646.163656  \\
7   1711.977507  \\
8   1525.565131  \\
9   1471.037321  \\
};
\end{axis}
\end{tikzpicture}%
 &\begin{tikzpicture}
\begin{axis}[%
width=\iwidth,
height=\iheight,
at={(0,0)},
scale only axis,
xmin=0,
xmax=10,
ymin=1000,
ymax=140000,
ymode=log,
yminorticks=true
]
\addplot [color=black,thick,solid,mark=*]
  table[row sep=crcr]{%
0   128192.351240 \\
1   17136.222529  \\
2   24942.525886  \\
3   13966.916857  \\
4   13471.226765  \\
5   7814.638377   \\
6   7479.520552   \\
7   6766.057862   \\
8   6490.149522   \\
9   5970.142726   \\
};
 \addplot [color=blue,thick,solid,mark=square*]
   table[row sep=crcr]{%
 0   41711.90412  \\
 1   9002.434694  \\
 2   8357.646375  \\
 3   5530.804004  \\
 4   7269.298591  \\
 5   4828.055263  \\
 6   5470.788912  \\
 7   4384.861119  \\
 8   5088.750380  \\
 9   3900.350273  \\
 };
\end{axis}
\end{tikzpicture}%
\\
		& \multicolumn{4}{c}{Projected Gauss-Newton-CG (PNCG) iterations}
		\end{tabular}
	\end{center}
	\caption{Convergence history for 3D Full Waveform Inversion (FWI) example. Columns represent the steps of the frequency continuation strategy. The first row shows the value of the objective function and the second row shows the norm of the projected gradient at each projected Gauss-Newton-CG iteration (black circles represent first cycle and blue squares represent second cycle if performed).}
	\label{fig:FWIhis}
\end{figure}

\RTHedit{The convergence history for the frequency continuation scheme is given in Fig.~\ref{fig:FWIhis} and  }the FWI reconstruction is visualized in Fig.~\ref{fig:mfwi}. \RTHedit{In each projected Gauss-Newton iteration, the maximum number of 9 projected PCG iterations is performed and the step size of one satisfies the Armijo line search condition. In the final iteration the bound constraints are active in 5,882 voxels (roughly 0.25\% of voxels).} As to be expected, given the low frequency data, it shows a blurred version of the true model. Including data that corresponds to higher frequencies will enable a sharper reconstruction, but will also require a finer mesh, which is much more expensive to process. This inversion took about three days of computations using the shared memory computer described above. We note that the data in this experiment corresponds to relatively low frequencies, which are often not available in real life experiments. In the absence of those frequencies, a good smooth initial model needs to be provided for the inversion. Generating the smooth model may be done using other techniques or inverse problems, like those shown in the next section, but is generally an open question and beyond the scope of this paper. For a more detailed discussion about handling FWI in the absence of low-frequency data, see \cite{treister2016full}.

\subsection{Joint travel time and DC-resistivity survey} 
\label{sub:3d_joint_travel_time_and_dc_resistivity_survey}
In this example, we demonstrate \jInv's potential for solving multiphysics inverse problems in one common framework and using different parallel computing schemes. \RTHedit{In reservoir exploration, e.g., different physical properties of rocks are commonly considered to identify the true rock type---this field is called \emph{petrophysics} \cite{schon2015physical,archie1950introduction}. Here we wish to illustrate such an integration in a joint inverse problem that includes both seismic and electromagnetic data.} To this end, we consider the problem of jointly \RTHedit{fitting} travel time and DC-resistivity data \RTHedit{for enhancing the quality of the reconstruction}.  We again consider a synthetic example for reconstructing the 3D SEG/EAGE model in Fig.~\ref{fig:mtrue}. This time, the domain is divided into a coarser $64\times 64\times 32$ equally sized mesh cells of size $211\,{\rm m}\times 211\,{\rm m} \times 131\,{\rm m}$ each.

\begin{figure}
\begin{center}
  \subfigure{\quad\quad
	\iwidth =35mm
	\iheight=25mm
%
%
\definecolor{mycolor1}{rgb}{0.00000,0.44700,0.74100}%
\begin{tikzpicture}[font=\scriptsize]

\begin{axis}[%
width=0.95092\iwidth,
height=\iheight,
at={(0\iwidth,0\iheight)},
scale only axis,
xmin=1.5,
xmax=5,
xlabel={wave velocity},
ymin=0,
ymax=1.5,
ylabel={conductivity},
ylabel style={at={(+.15,0.5)}},
xlabel style={at={(+.5,0.1)}}
]
\addplot [color=black,solid,forget plot,line width=1.5pt]
  table[row sep=crcr]{%
1.5	1.49999999999987\\
1.51	1.49666666666651\\
1.52	1.49333333333315\\
1.53	1.48999999999977\\
1.54	1.48666666666639\\
1.55	1.48333333333299\\
1.56	1.47999999999959\\
1.57	1.47666666666616\\
1.58	1.47333333333272\\
1.59	1.46999999999925\\
1.6	1.46666666666575\\
1.61	1.46333333333222\\
1.62	1.45999999999864\\
1.63	1.45666666666501\\
1.64	1.45333333333132\\
1.65	1.44999999999755\\
1.66	1.44666666666368\\
1.67	1.44333333332969\\
1.68	1.43999999999556\\
1.69	1.43666666666126\\
1.7	1.43333333332674\\
1.71	1.42999999999197\\
1.72	1.42666666665688\\
1.73	1.42333333332141\\
1.74	1.41999999998547\\
1.75	1.41666666664896\\
1.76	1.41333333331176\\
1.77	1.40999999997371\\
1.78	1.40666666663463\\
1.79	1.4033333332943\\
1.8	1.39999999995243\\
1.81	1.39666666660871\\
1.82	1.39333333326271\\
1.83	1.38999999991395\\
1.84	1.38666666656181\\
1.85	1.38333333320557\\
1.86	1.37999999984433\\
1.87	1.37666666647699\\
1.88	1.37333333310222\\
1.89	1.3699999997184\\
1.9	1.36666666632356\\
1.91	1.36333333291529\\
1.92	1.35999999949064\\
1.93	1.35666666604606\\
1.94	1.35333333257719\\
1.95	1.34999999907872\\
1.96	1.34666666554419\\
1.97	1.34333333196573\\
1.98	1.33999999833375\\
1.99	1.33666666463657\\
2	1.33333333085995\\
2.01	1.32999999698655\\
2.02	1.32666666299526\\
2.03	1.32333332886033\\
2.04	1.31999999455043\\
2.05	1.31666666002735\\
2.06	1.31333332524459\\
2.07	1.30999999014546\\
2.08	1.30666665466093\\
2.09	1.3033333187069\\
2.1	1.29999998218092\\
2.11	1.2966666449582\\
2.12	1.29333330688672\\
2.13	1.28999996778128\\
2.14	1.28666662741632\\
2.15	1.28333328551705\\
2.16	1.27999994174876\\
2.17	1.27666659570373\\
2.18	1.27333324688531\\
2.19	1.26999989468855\\
2.2	1.26666653837658\\
2.21	1.26333317705183\\
2.22	1.25999980962099\\
2.23	1.25666643475238\\
2.24	1.25333305082396\\
2.25	1.24999965585999\\
2.26	1.24666624745403\\
2.27	1.24333282267497\\
2.28	1.23999937795269\\
2.29	1.23666590893886\\
2.3	1.23333241033722\\
2.31	1.2299988756971\\
2.32	1.2266652971618\\
2.33	1.22333166516219\\
2.34	1.21999796804365\\
2.35	1.21666419161156\\
2.36	1.21333031857801\\
2.37	1.20999632788779\\
2.38	1.20666219389775\\
2.39	1.20332788537732\\
2.4	1.19999336429143\\
2.41	1.19665858431845\\
2.42	1.19332348904534\\
2.43	1.18998800976997\\
2.44	1.18665206282496\\
2.45	1.18331554631906\\
2.46	1.17997833616931\\
2.47	1.17664028126978\\
2.48	1.17330119760925\\
2.49	1.16996086110923\\
2.5	1.16661899890453\\
2.51	1.16327527872795\\
2.52	1.1599292959878\\
2.53	1.15658055803768\\
2.54	1.15322846503006\\
2.55	1.14987228661385\\
2.56	1.14651113357715\\
2.57	1.14314392334313\\
2.58	1.13976933799374\\
2.59	1.13638577321297\\
2.6	1.13299127620026\\
2.61	1.12958347019321\\
2.62	1.12615946274409\\
2.63	1.12271573430095\\
2.64	1.11924800293549\\
2.65	1.11575106021629\\
2.66	1.11221857222841\\
2.67	1.10864283856952\\
2.68	1.10501450079081\\
2.69	1.10132219018701\\
2.7	1.09755210307493\\
2.71	1.09368748975418\\
2.72	1.08970804127229\\
2.73	1.08558915602912\\
2.74	1.081301066343\\
2.75	1.07680780368216\\
2.76	1.07206598083913\\
2.77	1.06702337065764\\
2.78	1.06161726513852\\
2.79	1.05577260747938\\
2.8	1.04939990510306\\
2.81	1.04239295704746\\
2.82	1.03462646812767\\
2.83	1.02595367973844\\
2.84	1.016204228122\\
2.85	1.0051825498472\\
2.86	0.992667292969932\\
2.87	0.978412356630931\\
2.88	0.962150356081689\\
2.89	0.943599460318\\
2.9	0.922474615852763\\
2.91	0.898504062837247\\
2.92	0.871451654530413\\
2.93	0.841144694695788\\
2.94	0.807505751252098\\
2.95	0.77058526611312\\
2.96	0.730590060121716\\
2.97	0.687901582359249\\
2.98	0.643077686728548\\
2.99	0.5968334329065\\
3	0.55\\
3.01	0.503465571077375\\
3.02	0.458106565192801\\
3.03	0.414720231152815\\
3.04	0.373969327425347\\
3.05	0.33634649124578\\
3.06	0.302161140953866\\
3.07	0.271547028623672\\
3.08	0.244485227956015\\
3.09	0.220835979658126\\
3.1	0.200373208825909\\
3.11	0.1828170074001\\
3.12	0.167861209770749\\
3.13	0.155194846582288\\
3.14	0.144517476254563\\
3.15	0.13554912156682\\
3.16	0.12803586248951\\
3.17	0.121752182862318\\
3.18	0.116501056566029\\
3.19	0.112112578065825\\
3.2	0.10844174970149\\
3.21	0.10536586452727\\
3.22	0.102781781443551\\
3.23	0.100603280684731\\
3.24	0.0987586089148164\\
3.25	0.0971882686792017\\
3.26	0.0958430710286816\\
3.27	0.0946824477188109\\
3.28	0.0936730064246065\\
3.29	0.0927873058053213\\
3.3	0.0920028247568742\\
3.31	0.0913011002206299\\
3.32	0.0906670094017936\\
3.33	0.0900881744802936\\
3.34	0.0895544704208892\\
3.35	0.0890576190328819\\
3.36	0.0885908548363987\\
3.37	0.0881486504812604\\
3.38	0.0877264914034979\\
3.39	0.0873206910901841\\
3.4	0.0869282397684306\\
3.41	0.0865466805606241\\
3.42	0.0861740081801606\\
3.43	0.0858085861053935\\
3.44	0.0854490788883232\\
3.45	0.0850943968506295\\
3.46	0.0847436509124244\\
3.47	0.0843961157054778\\
3.48	0.0840511994571125\\
3.49	0.0837084194057829\\
3.5	0.0833673817348602\\
3.51	0.0830277651960159\\
3.52	0.0826893077450007\\
3.53	0.0823517956365375\\
3.54	0.0820150545264126\\
3.55	0.081678942211725\\
3.56	0.0813433427079833\\
3.57	0.0810081614170792\\
3.58	0.0806733211853673\\
3.59	0.080338759087988\\
3.6	0.0800044238057136\\
3.61	0.079670273485193\\
3.62	0.0793362739935591\\
3.63	0.0790023974947471\\
3.64	0.0786686212882514\\
3.65	0.0783349268619634\\
3.66	0.0780012991196366\\
3.67	0.0776677257507964\\
3.68	0.077334196716838\\
3.69	0.0770007038318946\\
3.7	0.0766672404210058\\
3.71	0.0763338010413333\\
3.72	0.0760003812548001\\
3.73	0.0756669774426705\\
3.74	0.0753335866543366\\
3.75	0.0750002064840032\\
3.76	0.0746668349701247\\
3.77	0.0743334705133993\\
3.78	0.0740001118098947\\
3.79	0.0736667577965165\\
3.8	0.0733334076065403\\
3.81	0.0730000605333519\\
3.82	0.0726667160008806\\
3.83	0.0723333735394901\\
3.84	0.0720000327663215\\
3.85	0.0716666933692639\\
3.86	0.0713333550938862\\
3.87	0.0710000177327814\\
3.88	0.0706666811168789\\
3.89	0.0703333451083607\\
3.9	0.0700000095948871\\
3.91	0.0696666744848881\\
3.92	0.0693333397037248\\
3.93	0.0690000051905602\\
3.94	0.0686666708958081\\
3.95	0.0683333367790531\\
3.96	0.0680000028073552\\
3.97	0.0676666689538677\\
3.98	0.0673333351967106\\
3.99	0.0670000015180517\\
4	0.0666666679033589\\
4.01	0.0663333343407887\\
4.02	0.0660000008206898\\
4.03	0.0656666673351971\\
4.04	0.0653333338779013\\
4.05	0.0650000004435798\\
4.06	0.0646666670279766\\
4.07	0.0643333336276241\\
4.08	0.0640000002396965\\
4.09	0.0636666668618914\\
4.1	0.063333333492333\\
4.11	0.0630000001294928\\
4.12	0.0626666667721255\\
4.13	0.0623333334192164\\
4.14	0.0620000000699391\\
4.15	0.0616666667236201\\
4.16	0.0613333333797108\\
4.17	0.0610000000377643\\
4.18	0.0606666666974165\\
4.19	0.0603333333583709\\
4.2	0.0600000000203857\\
4.21	0.0596666666832644\\
4.22	0.0593333333468465\\
4.23	0.0590000000110015\\
4.24	0.058666666675623\\
4.25	0.0583333333406245\\
4.26	0.0580000000059354\\
4.27	0.0576666666714983\\
4.28	0.0573333333372662\\
4.29	0.0570000000032012\\
4.3	0.0566666666692723\\
4.31	0.0563333333354541\\
4.32	0.056000000001726\\
4.33	0.0556666666680714\\
4.34	0.0553333333344766\\
4.35	0.0550000000009304\\
4.36	0.0546666666674238\\
4.37	0.0543333333339494\\
4.38	0.0540000000005013\\
4.39	0.0536666666670746\\
4.4	0.0533333333336652\\
4.41	0.05300000000027\\
4.42	0.0526666666668863\\
4.43	0.0523333333335121\\
4.44	0.0520000000001454\\
4.45	0.0516666666667849\\
4.46	0.0513333333334296\\
4.47	0.0510000000000783\\
4.48	0.0506666666667304\\
4.49	0.0503333333333851\\
4.5	0.0500000000000421\\
4.51	0.0496666666667009\\
4.52	0.0493333333333612\\
4.53	0.0490000000000227\\
4.54	0.0486666666666851\\
4.55	0.0483333333333483\\
4.56	0.0480000000000122\\
4.57	0.0476666666666766\\
4.58	0.0473333333333414\\
4.59	0.0470000000000066\\
4.6	0.046666666666672\\
4.61	0.0463333333333377\\
4.62	0.0460000000000035\\
4.63	0.0456666666666695\\
4.64	0.0453333333333357\\
4.65	0.0450000000000019\\
4.66	0.0446666666666682\\
4.67	0.0443333333333346\\
4.68	0.044000000000001\\
4.69	0.0436666666666675\\
4.7	0.043333333333334\\
4.71	0.0430000000000005\\
4.72	0.0426666666666671\\
4.73	0.0423333333333337\\
4.74	0.0420000000000003\\
4.75	0.0416666666666669\\
4.76	0.0413333333333335\\
4.77	0.0410000000000002\\
4.78	0.0406666666666668\\
4.79	0.0403333333333334\\
4.8	0.0400000000000001\\
};
\end{axis}
\end{tikzpicture}%
\label{fig:vcmodel}
}
\end{center}
\caption{\footnotesize The mapping from velocity to conductivity given by~\eqref{eq:velToCond}. }
\label{fig:SEG_setting}
\end{figure}
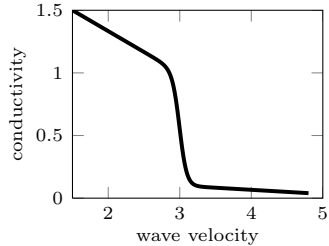

\paragraph{Experimental setting}
For the DC resistivity problem we use $16$ dipole sources in both the $x_1$ and $x_2$ directions. The sources are placed on the top surface of the 3D volume and arranged on a $4\times 4$ grid localized in the center of the domain. The spacing between the sources is around $840\,{\rm m}$, and the length of each source is around $10.1\,{\rm km}$. For each source, the resulting potential field is measured using dipole receivers placed on the top surface with of approximate length $840\,{\rm m}$. In both the $x_1$ and $x_2$ direction 841 receivers are placed on a $29 \times 29$ grid. Generally, the relation between the wave velocity and the ground conductivity may vary from site to site. Here, we assume a relation that generates a conductivity ratio contrast of about 10 between the conductivity of the salt body and the rest of the ground. Mathematically we use the function
\begin{equation}\label{eq:velToCond}
\textstyle\sigma(m) = \left(2-\frac{m}{c}\right)\left(\frac{b-a}{2}\cdot(\tanh(10\cdot(c-m))+1)+a\right),
\end{equation}
where $m$ is the velocity, $a$ and $b$ are the conductivity values, set to 0.1 and 1.0 respectively, and $c=3.0$ is the velocity in which the contrast is centered. $\sigma(m)$ is also illustrated in Fig.~\ref{fig:vcmodel}. To accelerate computations, a coarser mesh consisting of $32\times 32 \times 16$ cells of size around $420\,{\rm m} \times 420\,{\rm m} \times 260\,{\rm m}$ is used to simulate the fields. To generate the data $\bfd \in \R^{1682 \times 32}$ we use the true model and $\sigma(m)$ in~\eqref{eq:velToCond}, and add Gaussian white i.i.d noise with standard deviation $0.01\cdot \bar{\bfd}$, where $\bar{\bfd}$ denotes the mean of the true data.

For the travel time tomography we use 36 sources that are spaced equidistantly $2.11\,{\rm km}$ apart from each other on the top surface, and arranged on a $6 \times 6$ grid. The travel time is recorded using $3,481$ receivers on the top surface that are placed 211 m of each other on a $59\times 59$ grid. The same $64 \times 64 \times 32$ mesh as for the inversion is used to simulate the data, which is done using the fast marching method for the true velocity model (converted to squared slowness). Following that, white Gaussian i.i.d noise with standard deviation of $0.01\cdot \bar{\bfd}$ is added to the data.

\paragraph{Inversion parameters and parallelization settings}
For all the experiments we used 10 iterations of the projected Gauss-Newton with at most 8 projected CG iterations in each step, starting from the same linear reference model shown in Fig.~\ref{fig:mref}. Similar to the FWI experiment, we enforce a smooth reconstruction by a diffusion regularizer. Again, we choose a very small regularization parameter and use the Hessian of the regularizer as a preconditioner. We also use the same lower and upper bounds on the velocity model to keep the model physical.

Because of the different nature of the problems, we use different parallelization techniques for solving the two discretized PDEs. To solve the linear system~\eqref{dc}, we apply MUMPS using a shared-memory parallelization on a single Julia worker. We store the factorization in each Gauss-Newton iteration to speed up the sensitivity matrix-vector products. All DC sources are treated on that worker, and the forward problems are solved on a coarse $32\times32\times16$ mesh.
For the travel time inversion, equal batches of sources are distributed among all available workers (8), and all batches of forward problems are solved (one-by-one) in parallel using a $64\times64\times32$ mesh. The parallelization setting is as such because of the sequential nature (and code) of the Fast Marching method; see~\cite{TreisterHaber2016} for details. While the setting for each of those problems separately is simple in principle, the setting of the joint problem is more complicated, since these are completely different parallelization scenarios. Thanks to the abstraction in \jInv, setting this up requires only to add all misfit problems into one array. A similar setting was also used in the joint inversion of FWI and travel time tomography in \cite{treister2016full}.

\begin{figure}
\begin{center}
	\newcommand{\image}[1]{\includegraphics[width=0.32\linewidth,trim=75 140 130 120,clip=true]{#1}}
  \subfigure[\footnotesize DC resistivity inversion]{\image{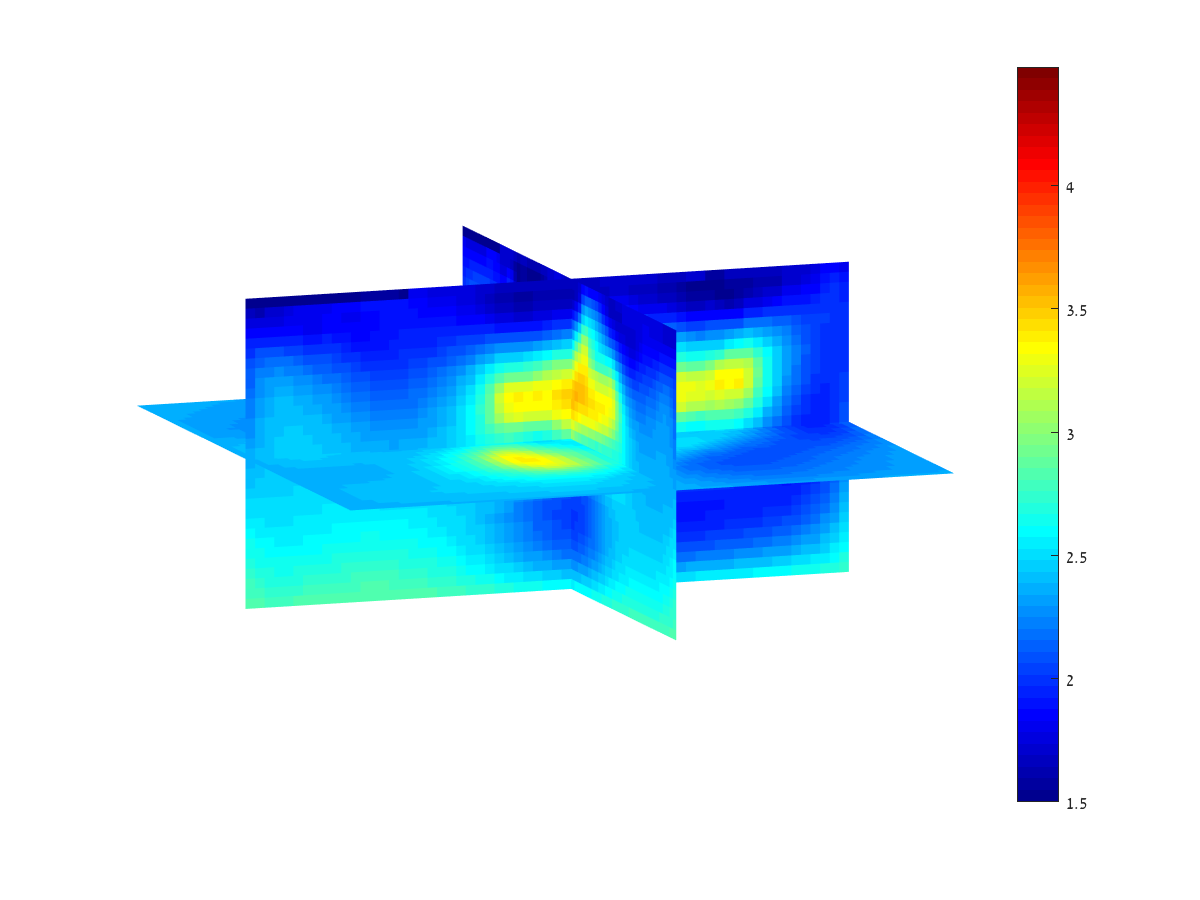}\label{fig:mDC}}
  \subfigure[\footnotesize Travel time inversion]{\image{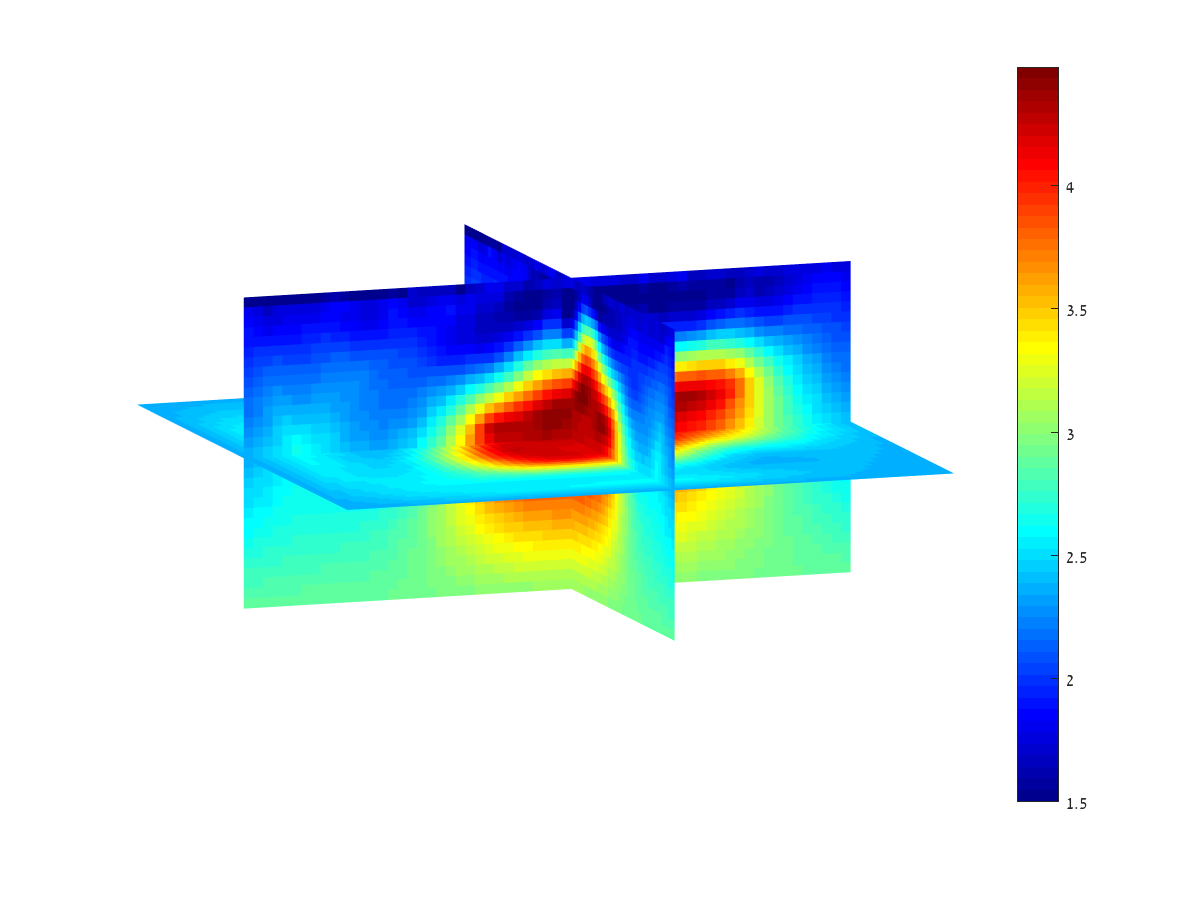}\label{fig:mEik}}
  \subfigure[\footnotesize Joint DC resistivity and travel time inversion]{\image{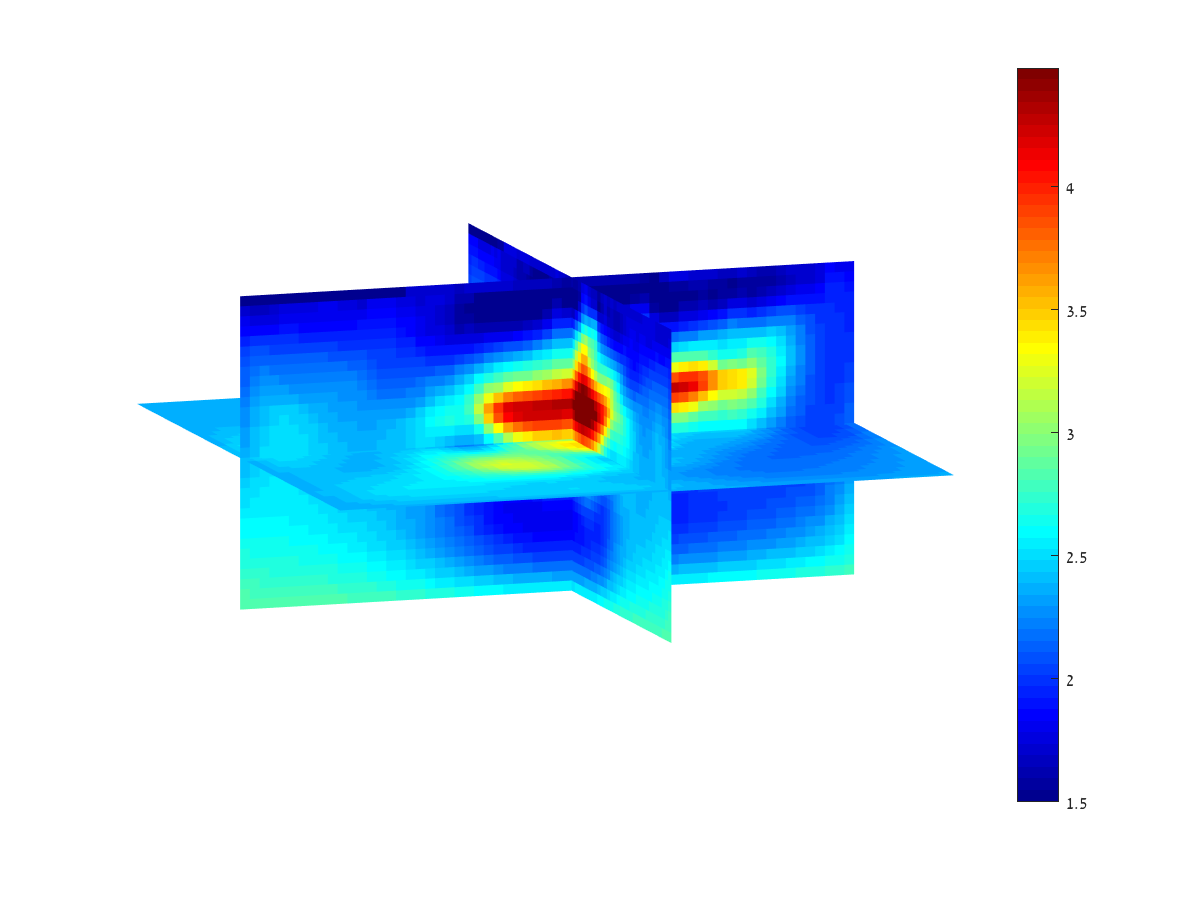}\label{fig:mJoint}}
\end{center}
\caption{\footnotesize Results of 3D multiphysics inversion using DC resistivity and travel time tomography data. Starting guess (top left) and reconstructions of the model for different data is visualized using orthogonal slices. Using the DC resistivity data, the reconstruction captures the overall shape of the salt reservoir, but intensity values are inaccurate; compare Fig~\ref{fig:mDC} to Fig.~\ref{fig:mtrue}. Travel time tomography provides more accurate estimate of intensity, but does not provide spatial localization; see~Fig.~\ref{fig:mEik}. The joint reconstruction (bottom right) combines the advantages of both modalities. }
\label{fig:JointInv}
\end{figure}

\paragraph{Reconstructions}

\begin{figure}[t]
	\begin{center}
		\newcommand{\rottext}[1]{\rotatebox{90}{\hbox to 30mm{\hss #1\hss}}}
		\iwidth=20mm
		\iheight=20mm
		\footnotesize
		\begin{tabular}{cccc}
			& DC Resistivity & Travel Time Tomography & Joint Inversion \\
			\rottext{obj. function}	
				& \begin{tikzpicture}
\begin{axis}[%
width=\iwidth,
height=\iheight,
at={(0,0)},
scale only axis,
xmin=0,
xmax=10,
ymin=1,
ymax=100,
ymode=log,
yminorticks=true
]
\addplot [color=black,thick,solid]
  table[row sep=crcr]{%
0    101.91740 \\
1     90.85330 \\
2     79.85700 \\
3     61.2376 \\
4     35.8073 \\
5     23.6143 \\
6     15.9046 \\
7     11.9587 \\
8      9.8067 \\
9      8.4508 \\
10     7.6897 \\
};
\end{axis}
\end{tikzpicture}%
 & \begin{tikzpicture}
\begin{axis}[%
width=\iwidth,
height=\iheight,
at={(0,0)},
scale only axis,
xmin=0,
xmax=10,
ymin=.1,
ymax=301,
ymode=log,
yminorticks=true
]
\addplot [color=black,thick,solid]
  table[row sep=crcr]{%
0    300.8040 \\
1    116.5396 \\
2     48.5225 \\
3     18.3869 \\
4      5.4353 \\
5      1.7342 \\
6      1.0913 \\
7      0.9933 \\
8      0.9626 \\
9      0.9441 \\
10     0.9310 \\
};
\end{axis}
\end{tikzpicture}%
 & \begin{tikzpicture}
\begin{axis}[%
width=\iwidth,
height=\iheight,
at={(0,0)},
scale only axis,
xmin=0,
xmax=10,
ymin=1,
ymax=360,
ymode=log,
yminorticks=true
]
\addplot [color=black,thick,solid]
  table[row sep=crcr]{%
0   351.7381 \\
1   272.8929 \\
2   205.3458 \\
3   149.6147 \\
4    67.5661 \\
5    32.5469 \\
6    17.0471 \\
7     9.6174 \\
8     6.2751 \\
9     5.3704 \\
10    5.1198 \\
};
\end{axis}
\end{tikzpicture}%
\\
			\rottext{opt. condition}	
				& \begin{tikzpicture}
\begin{axis}[%
width=\iwidth,
height=\iheight,
at={(0,0)},
scale only axis,
xmin=0,
xmax=10,
ymin=0.1,
ymax=10,
ymode=log,
yminorticks=true
]
\addplot [color=black,thick,solid]
  table[row sep=crcr]{%
0     2.93900 \\
1     2.77300 \\
2     2.61020 \\
3     4.9309 \\
4     4.5092 \\
5     3.7592 \\
6     2.6776 \\
7     1.8376 \\
8     1.2389 \\
9     0.7250 \\
};
\end{axis}
\end{tikzpicture}%
 & \begin{tikzpicture}
\begin{axis}[%
width=\iwidth,
height=\iheight,
at={(0,0)},
scale only axis,
xmin=0,
xmax=10,
ymin=0,
ymax=10,
ymode=log,
yminorticks=true
]
\addplot [color=black,thick,solid]
  table[row sep=crcr]{%
0    8.91790 \\
1    4.78000 \\
2    2.88420 \\
3    1.6499 \\
4    0.7541 \\
5    0.2425 \\
6    0.0620 \\
7    0.0456 \\
8    0.0299 \\
9    0.0293 \\
};
\end{axis}
\end{tikzpicture}%
 & \begin{tikzpicture}
\begin{axis}[%
width=\iwidth,
height=\iheight,
at={(0,0)},
scale only axis,
xmin=0,
xmax=10,
ymin=.1,
ymax=10,
ymode=log,
yminorticks=true
]
\addplot [color=black,thick,solid]
  table[row sep=crcr]{%
0  8.8313 \\
1  7.4189 \\
2  6.2957 \\
3  6.7203 \\
4  3.7965 \\
5  2.2478 \\
6  1.3099 \\
7  0.7141 \\
8  0.3069 \\
9  0.2319 \\
};
\end{axis}
\end{tikzpicture}%
\\
				& \multicolumn{3}{c}{projected Gauss-Newton-CG iterations}
		\end{tabular}
	\end{center}
	\caption{\RTHedit{Convergence history for the DC Resistivity, travel time tomography, and joint inversion example (column-wise). For each iteration of the projected Gauss-Newton method, the value of the objective function (first row) and the norm of the projected gradient (second row) are shown.}}
	\label{fig:JointInvHis}
\end{figure}
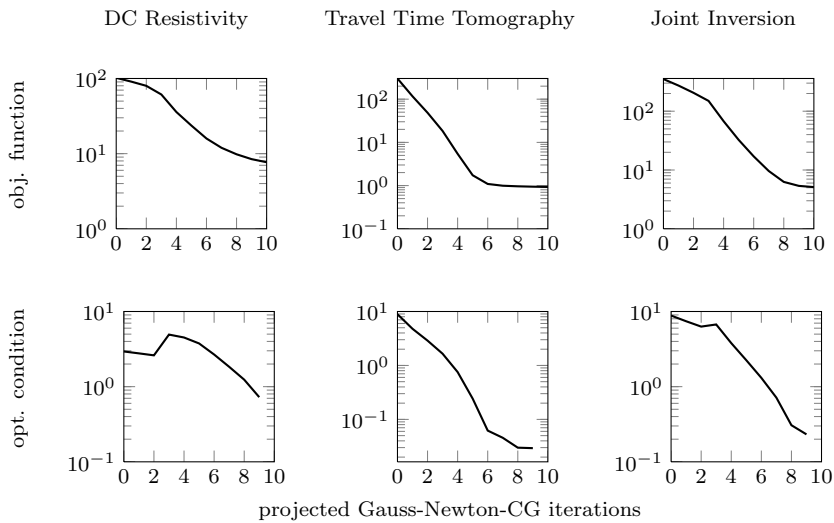

In all the experiments, the reconstructed model fits the observed data that is taken into account in the inversion up to the noise level.
\RTHedit{In all cases the projected Gauss-Newton method reduces the objective function value by at least one order of magnitude as can be seen in the convergence history in Fig.~\ref{fig:JointInvHis}. For all methods, exactly one line search step per iteration is needed and the number of projected PCG iterations is between 3 and 10 (average is around 8).}
 The inversion result using only the DC resistivity data is shown in Fig.~\ref{fig:mDC}.  It can be seen that the DC data contains information about the location of the salt dome, however, there is a significant error in the value of the reconstructed velocity. The reconstruction using only the travel time data is shown in Fig.~\ref{fig:mEik}. It can be seen that the reconstruction contains accurate information about the location of the top surface of the salt dome as well as the absolute value of the velocity. However, the reconstruction does not provide accurate information about the bottom of the salt dome, which is smoothed due to the regularization. By jointly inverting the two problems we aim to combine the benefits of both modalities, and indeed this is achieved in our experiment whose result appears in Fig.~\ref{fig:mJoint}. The shape of the salt body is well-captured thanks for the DC data, and the value for the velocity is more accurate than in the DC experiment, thanks for the travel time experiment. In terms of timings, the DC inversion alone took about 4 minutes, the travel time inversion took about 11 minutes, and the joint inversion took about 13 minutes. These experiments were performed using the laptop whose specification was mentioned earlier.

\subsection{Weak scaling tests for forward problems} 
\label{sub:weak_scaling_test_for_forward_problems}

Practical solutions of inverse problems need to consider problems where the number of sources frequencies and receivers grow. Therefore, a relevant  scalability test is to examine the behavior of the code as the
number of forward problems increase. Here we  evaluate the parallel efficiency of the static scheduling approach to computing forward problem described in Sec.~\ref{sub:static} by means of a weak scaling test. In this test we time the evaluation of the forward problems for different numbers of workers, while keeping the workload per worker constant. The parallel efficiency is computed as
\begin{equation}\label{eq:efficiency}
	{\rm efficiency}(n) = \frac{t(1)}{t(n)} \cdot 100,
\end{equation}
where $t(n)$ is the runtime measured using $n$ workers.

We consider the travel time tomography and DC resistivity problem described in the previous section. As mentioned above, \jInv\ supports various platforms and settings. Here we give two examples on a shared memory computer and a cloud computing engine.

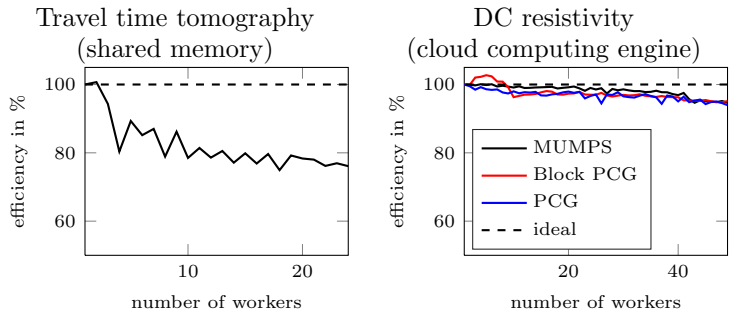
\begin{figure}[t]
	\begin{center}
		\begin{tabular}{cc}
			Travel time tomography  & DC resistivity\\
           (shared memory) &  (cloud computing engine)\\
			\begin{tikzpicture}[font=\scriptsize]
\begin{axis}[%
width=35mm,
height=25mm,
at={(0,0)},
scale only axis,
xmin=1,
xmax=24,
ymin=50,
ymax=105,
xlabel={number of workers},
ylabel={efficiency in \%},
yminorticks=true,
ylabel style={at={(+.1,0.5)}},
xlabel style={at={(+.5,0.05)}}
]
\addplot [color=black,thick,solid]
  table[row sep=crcr]{%
1    100.0\\
2    100.7\\
3    94.31\\
4    80.38\\
5    89.32\\
6    85.17\\
7    87.01\\
8    78.90\\
9    86.19\\
10   78.51\\
11   81.42\\
12   78.6\\
13   80.54\\
14   77.12\\
15   79.86\\
16   76.87\\
17   79.65\\
18   74.94\\
19   79.23\\
20   78.34\\
21   78.01\\
22   76.17\\
23   76.95\\
24   76.1\\
};

\addplot [color=black,thick,dashed]
  table[row sep=crcr]{%
1    100.0  \\
2    100.0  \\
3    100.0  \\
4    100.0  \\
5    100.0  \\
6    100.0  \\
7    100.0  \\
8    100.0  \\
9    100.0  \\
10   100.0  \\
11   100.0  \\
12   100.0  \\
13   100.0  \\
14   100.0  \\
15   100.0  \\
16   100.0  \\
17   100.0  \\
18   100.0  \\
19   100.0  \\
20   100.0  \\
21   100.0  \\
22   100.0  \\
23   100.0  \\
24   100.0  \\
};

\end{axis}
\end{tikzpicture}%
			&
			\begin{tikzpicture}[font=\scriptsize]
\begin{axis}[%
width=35mm,
height=25mm,
at={(0,0)},
scale only axis,
xmin=1,
xmax=49,
ymin=50,
ymax=105,
xlabel={number of workers},
ylabel={efficiency in \%},
yminorticks=true,
legend style={at={(0.03,0.03)},anchor=south west,legend cell align=left,align=left,draw=white!15!black},
ylabel style={at={(+.1,0.5)}},
xlabel style={at={(+.5,0.05)}}
]
\addplot [color=black,thick,solid]
  table[row sep=crcr]{%
49      94.66128088       \\
48      95.16755829       \\
47      94.83853631       \\
46      94.97907506       \\
45      95.07463489       \\
44      95.36064562       \\
43      94.65581687       \\
42      95.70883571       \\
41      97.53311269       \\
40      96.89092179       \\
39      97.19676293       \\
38      97.71831453       \\
37      97.74708272       \\
36      98.18222441       \\
35      97.85328084       \\
34      97.78902227       \\
33      98.07802976       \\
32      98.05864749       \\
31      98.27514761       \\
30      98.58185353       \\
29      98.31552888       \\
28      98.69429276       \\
27      97.37236753       \\
26      98.91157473       \\
25      98.54515288       \\
24      98.99261912       \\
23      98.14954578       \\
22      98.86888508       \\
21      99.41894896       \\
20      99.18384959       \\
19      99.03526984       \\
18      98.84039733       \\
17      99.31673537       \\
16      99.24878291       \\
15      99.18363012       \\
14      99.13288491       \\
13      99.02972681       \\
12      98.98045035       \\
11      99.46446931       \\
10      99.12038938       \\
9       99.38625005       \\
8       99.63862483       \\
7       99.41674386       \\
6       100.077699        \\
5       99.84748401       \\
4       100.0855948       \\
3       99.73676774       \\
2       99.93088688       \\
1       100               \\
};
\addlegendentry{MUMPS};

\addplot [color=red,thick,solid]
  table[row sep=crcr]{%
49     95.14334876  \\
48     94.79251549  \\
47     94.84707894  \\
46     95.09088829  \\
45     95.45846897  \\
44     95.23667314  \\
43     95.29899117  \\
42     95.28684012  \\
41     95.38118413  \\
40     96.25842109  \\
39     96.28159564  \\
38     96.34917865  \\
37     96.62336441  \\
36     96.36762685  \\
35     96.51483764  \\
34     96.70659465  \\
33     97.13689269  \\
32     96.86524193  \\
31     96.79313413  \\
30     96.96794944  \\
29     96.74529054  \\
28     96.48538468  \\
27     96.79402056  \\
26     97.53849277  \\
25     96.98688966  \\
24     97.00481977  \\
23     97.07813509  \\
22     97.84705479  \\
21     97.39019632  \\
20     97.31730512  \\
19     97.02670094  \\
18     97.07660655  \\
17     97.9013058   \\
16     98.08288115  \\
15     97.71369026  \\
14     97.23557661  \\
13     97.07482333  \\
12     97.03013667  \\
11     96.66047825  \\
10     96.30014331  \\
9      99.20884803  \\
8      100.8312107  \\
7      100.8683275  \\
6      102.386511   \\
5      102.7228537  \\
4      102.2378098  \\
3      102.0368815  \\
2      99.87068139  \\
1      100          \\
};
\addlegendentry{Block PCG};

\addplot [color=blue,thick,solid]
  table[row sep=crcr]{%
49     93.96295634 \\
48     94.55469285 \\
47     95.03269261 \\
46     94.86945098 \\
45     94.26246825 \\
44     95.58961454 \\
43     95.26097257 \\
42     94.83466537 \\
41     96.41807883 \\
40     95.06309526 \\
39     96.49039819 \\
38     96.77093415 \\
37     94.34894521 \\
36     96.23950984 \\
35     96.64801106 \\
34     97.1856317  \\
33     96.72891136 \\
32     96.16956989 \\
31     96.30647415 \\
30     96.54496139 \\
29     97.74213852 \\
28     96.93493707 \\
27     97.1037744  \\
26     94.47031611 \\
25     97.09387909 \\
24     96.63694369 \\
23     95.93380381 \\
22     97.84387149 \\
21     97.56411442 \\
20     97.9295793  \\
19     97.69442098 \\
18     97.26653806 \\
17     97.11335244 \\
16     96.74861886 \\
15     96.8089953  \\
14     97.83888028 \\
13     97.64307004 \\
12     97.75236053 \\
11     97.43344791 \\
10     97.99338982 \\
9      97.37883923 \\
8      97.67077346 \\
7      98.55618724 \\
6      98.42540991 \\
5      98.61765615 \\
4      99.22672083 \\
3      98.48356327 \\
2      99.49820653 \\
1      100         \\
};
\addlegendentry{PCG};

\addplot [color=black,thick,dashed]
  table[row sep=crcr]{%
49     100 \\
48     100 \\
47     100 \\
46     100 \\
45     100 \\
44     100 \\
43     100 \\
42     100 \\
41     100 \\
40     100 \\
39     100 \\
38     100 \\
37     100 \\
36     100 \\
35     100 \\
34     100 \\
33     100 \\
32     100 \\
31     100 \\
30     100 \\
29     100 \\
28     100 \\
27     100 \\
26     100 \\
25     100 \\
24     100 \\
23     100 \\
22     100 \\
21     100 \\
20     100 \\
19     100 \\
18     100 \\
17     100 \\
16     100 \\
15     100 \\
14     100 \\
13     100 \\
12     100 \\
11     100 \\
10     100 \\
9      100 \\
8      100 \\
7      100 \\
6      100 \\
5      100 \\
4      100 \\
3      100 \\
2      100 \\
1      100 \\
};
\addlegendentry{ideal};
\end{axis}
\end{tikzpicture}%
		\end{tabular}
	\end{center}
	\caption{Weak scaling tests for travel time tomography (left) and DC resistivity (right) forward problem. For the travel time tomography a shared memory machine with $x$ physical and $y$ virtual processors is used. For the DC resistivity problem a cloud computing engine is used. The ideal efficiency of 100\% is indicated by a dashed line. }
	\label{fig:getData}
\end{figure}

On the shared memory machine, we use the travel time tomography example since it is purely written in Julia and features a good ratio between memory access and computations.
We use a mesh size of $ 128  \times 128 \times 64$ to discretize the forward problem and solve a batch of 8 sources per worker.
We increase the number of workers from 1 to $24$ and measure the efficiency using~\eqref{eq:efficiency}.
The same batch of sources is solved on all workers in this experiment and so the largest experiment involves 192 sources.
The experiment is repeated five times and the average runtime is shown in the left subplot of Fig.~\ref{fig:getData}. The ideal efficiency of 100\%, corresponding to constant runtime, is indicated by a dashed line. The shortest average runtime of $20.7$ sec is observed using two workers and 16 sources. When increasing the number of workers, the runtime increases up to 27.89 seconds which corresponds to a weak scaling efficiency of around 74\%. This means that the time per source decreases from around 2.5 seconds using 1 worker to 0.14 seconds using 24 workers.

Scalability on the cloud computing engine is demonstrated using the DC resistivity problem.
Here, we use a relatively coarse mesh with $48\times48\times24$ cells for discretizing the forward problems in order to speed up computations. We use the same 10 dipole sources per worker. One main process is used to handle the communication and the number of workers is increased from 1 to 49.
We test two iterative solvers, Jacobi PCG and block Jacobi PCG, as well as the direct solver MUMPS.
As before, the parallel efficiency is computing using~\eqref{eq:efficiency} and plotted for a growing number of workers in the right subfigure of Fig.~\ref{fig:getData}.
The experiment is repeated five times and the average runtime is reported.
An almost perfect weak scaling is observed and the minimum observed efficiency across all experiments is around 93\%.

\subsection{Strong scaling test for iterative linear solver} 
\label{sub:strongScaling}

In some cases, one requires a large mesh for discretizing the forward problems, and solve for many sources
on this mesh. For such problems direct methods can be prohibitively expensive
or memory consuming and iterative methods become attractive. Here we test the behavior of
our package for the solution of a problem on a large mesh where the number of sources (right hand sides)
is fixed and the number of cores increases.
We thus demonstrate the strong scalability of the forward problem computations.
As an example, we consider the block conjugate gradient solver preconditioned with smoothed aggregation algebraic multigrid for the DC resistivity problem.
We use a mesh with $256\times 256 \times 128$ cells and we use 72 sources. As preconditioner, we apply a V(2,2) cycle with four levels, and use a shared memory version of MUMPS as a coarsest grid solver (the coarsest grid operator has 1723 unknowns).

Using the shared memory machine, we increase the number of threads assigned to Julia from 1 to 16, solve all forward problems, and measure the runtime.
The runtime decreases from around $1,995$ seconds using one thread to $358$ seconds using 16 threads, which is approximately a $5.5$x speedup. More detailed runtimes can be seen in Table~\ref{tab:strongScaling}.

\begin{table}
	\scriptsize
	\begin{center}
		\begin{tabular}{|c|c|c|}
			\hline
		number of threads & average runtime & speedup \\ \hline
		1                 &  1,994.3        &  1           \\
		2                 &  1,379.3        &  1.45           \\
		4                 &  826.3          &  2.41          \\
		8                 &  532.0          &  3.74           \\
		12                &   387.9         &  5.14           \\
		16                &   357.9         &  5.57        \\ \hline
		\end{tabular}
	\end{center}
	\caption{Strong scaling test for multigrid preconditioned block PCG solver using a 3D DC resistivity example with 72 sources and a regular mesh with $256\times256\times128$ cells. }\label{tab:strongScaling}
\end{table}

\section{Conclusions}
\label{sec6}
This paper presents \jInv, a new open source software package for PDE parameter estimation written in Julia.  At its core lie meshing tools that can be used to rapidly prototype inversion algorithm and experiment with different meshes and wrappers to efficient direct and iterative linear solvers. Further, it provides several options for misfit functions, regularizers and state-of-the-art methods for numerical optimization. \RTHedit{Another key feature of \jInv{} is the built-in parallelization that allow the user to easily solve inverse problems with many sources or frequencies in parallel.} \jInv\ is highly modular and thus can be easily extended to solve new applications or develop and compare inversion methods. Being written in a high-level language, \jInv{}'s source code is easy to read, modify, and extend. Due to Julia's Just-In-Time compiler we found that \jInv\ performs similar to highly optimized low-level code.

A main feature of \jInv\ are the provided tools to parallelize the inversion that can be applied across different computational architectures. It may seem that the problem~\eqref{genOpt} is embarrassingly simple to solve in parallel because it involves a sum of functions. However, a naive approach for parallelizing this solution typically requires significant data transfer between the workers. Each term of the misfit involves discretized PDE operators, sources, receivers, and gives rise to temporary results such as fields, preconditioners, or matrix factorizations that are often prohibitively expensive to communicate or even store on a single worker. Therefore, we present two parallelization strategies using dynamic and static scheduling. Depending on the computational architecture, both schemes have their merits. Dynamic scheduling is attractive when communication costs are low (due to good bandwidth or light-weight problem description) whereas the static assignment keeps communication costs to a minimum and balances the memory requirement across the workers. Weak and strong scaling tests show the effectiveness of the parallelization on different computational architectures.

\jInv\ is easy to install and the code is tested and used across different operating systems and computational platforms. The code is divided into several modules that can be combined and extended to tackle real-world problems.
For example, it is relatively straightforward to use jInv’s parallelization methods to solve problems not covered in this paper. To this end, the user is only required to provide a data structure defining the forward problem, a solver for the forward problems as well as methods for computing matrix vector products with the sensitivity matrices. Parallelizing over sources or frequencies is then possible, e.g., by stacking a number of problems into a vector. Therefore, we hope that our work will benefit other applications of PDE parameter estimation problems with many frequencies or sources such as hyper spectral DOT~\cite{SaibabaEtAl2015}. \RTHedit{On our website, we provide two tutorials that explain how to construct a new forward problem and use the parallelization.}

Another benefit of developing a common framework for inverting data of different modalities are simplifications of joint reconstructions that aim at extracting complementary information about the model.
While multiphysics reconstructions seem straightforward in theory, they are typically difficult to implement efficiently in practice and thus rarely used. Using the abstraction provided in the Julia language, jInv\ simplifies this process. We demonstrate this using a parallel multiphysics inversion using  a 3D example from geophysical imaging. In our example combining data from DC resistivity and travel time tomography improves the reconstruction quality considerably.

Our research is fully reproducible. jInv\ is made available freely under the MIT license and can be obtained at
\begin{center}
\href{https://github.com/JuliaInv/jInv.jl}{\texttt{https://github.com/JuliaInv/jInv.jl}}
\end{center}
Also the drivers used for the scalability tests as well as the 3D inversions are provided as examples. Being public domain, we anticipate active development and improvement of jInv\ in the future. Apart from providing more examples from different modalities a major focus will be on algorithmic improvements.

\section{Acknowledgements}
\RTHedit{The authors would like to thank Patrick Belliveau, UBC, for his great help in the development of this package and Jiahao Chen, MIT, for providing invaluable insight into the Julia language.}
The research leading to these results has received funding from the European Union's - Seventh Framework Programme (FP7/2007-2013) under grant agreement no 623212 - MC Multiscale Inversion. LR is supported by  National Science Foundation (NSF) grant DMS 1522599.

\bibliographystyle{abbrv}
\bibliography{jInv}

\end{document}